%% file: main.tex
\documentclass[12pt]{iopart}
\usepackage{iopams}

\usepackage{graphicx} 
\usepackage{subcaption}    
\usepackage{geometry}
\usepackage{enumitem}   
\usepackage{booktabs}
\usepackage{longtable}
\usepackage{ltcaption} 
\usepackage{lscape}
\usepackage{titlesec}
\usepackage{bold-extra}

\usepackage{tabularx} 
\usepackage{booktabs} 
\usepackage[none]{hyphenat} 
\usepackage{hyperref}
\usepackage{multirow}
\usepackage{xcolor} 
	
\usepackage{float} 
\usepackage{placeins} 

\begin{document}
\maketitle


\title[Decoding Saccadic Eye Movements from Brain Signals Using an Endovascular Neural Interface]{Decoding Saccadic Eye Movements from Brain Signals Using an Endovascular Neural Interface}

\author{Suleman Rasheed$^{1,2,3}$, James Bennett$^{3, 4}$, Peter E. Yoo$^{3, 4}$, Anthony N. Burkitt$^{1, 2}$, David B. Grayden$^{1, 2}$}

\address{$^1$ Department of Biomedical Engineering, The University of Melbourne, Australia}
\address{$^2$ Graeme Clark Institute of Biomedical Engineering, The University of Melbourne, Australia}
\address{$^3$ Synchron Inc., Brooklyn, New York, United States of America}
\address{$^4$ Vascular Bionics Laboratory, Department of Medicine, Royal Melbourne Hospital, The University of Melbourne, Australia}

\ead{sulemanr@student.unimelb.edu.au}

\input{0_Abstract}

\maketitle

\input{1_Introduction}

\input{2_Methods}

\input{3_Results}

\input{4_Discussion}

\input{5_Conclusion}

\input{Acknowledgment}

\newpage
\input{6_Appendices}

\newpage
\section*{References}
\bibliographystyle{unsrt}
\bibliography{references}


\end{document}

%% file: 0_Abstract.tex


\begin{abstract}

    \noindent
    \textit{Objective.}
    An Oculomotor Brain-Computer Interface~(BCI) records neural activity from brain regions involved in planning eye movements and translates this activity into control commands. 
    While previous successful studies have relied on invasive implants in non-human primates or electrooculography~(EOG) artefacts in human electroencephalogram (EEG) data, this study aimed to demonstrate the feasibility of an oculomotor BCI using a minimally invasive endovascular Stentrode\textsuperscript{TM} device implanted near the supplementary motor area of a patient with Amyotrophic Lateral Sclerosis~(ALS).

    \noindent
    \textit{Approach.}
    One participant performed self-paced visually-guided and free-viewing saccade tasks in four directions (left, right, up, down) while endovascular EEG and eye gaze recordings were collected. Visually-guided saccades were cued with visual stimuli, whereas free-viewing saccades were self-directed without explicit cues.     
    Brain signals were pre-processed to remove cardiac artefacts, downsampled, and classified using a Random Forest algorithm. 
    For saccade onset classification (fixation vs.\ saccade), features in time and frequency domains were extracted after xDAWN denoising, while for saccade direction classification, the downsampled time series were classified directly without explicit feature extraction.

    \noindent
    \textit{Main results.}
    The neural responses of visually-guided saccades overlapped with cue-evoked potentials, while free-viewing saccades exhibited saccade-related potentials that began shortly before eye movement, peaked approximately 50~ms after saccade onset, and persisted for around 200~ms. In the frequency domain, these responses appeared as a low-frequency synchronisation below 15~Hz. 
    Saccade onset classification was robust, achieving mean area under the receiver operating characteristic curve (AUC) scores of 0.88 within sessions and 0.86 across sessions. Saccade direction decoding yielded within-session AUC scores of 0.67 for four-class decoding and up to 0.75 for the best performing binary comparisons  (left vs.\ up and left vs.\ down).

    \noindent
    \textit{Significance.}
    This proof-of-concept study demonstrates the feasibility of an endovascular oculomotor BCI in a patient with ALS, establishing a foundation for future oculomotor BCI studies in human subjects.

\end{abstract}
\vspace{2pc}
\noindent{\it Keywords}: Oculomotor BCI, BMI, Intravascular, Endovascular, Stentrode, ALS, Saccade

\submitto{\JNE}
\maketitle



%% file: 1_Introduction.tex



\section{Introduction}
\label{Introduction}
A Brain-Computer Interface (BCI) acquires brain signals, analyses them, and translates them into commands that can control an output device, such as a computer cursor or robotic arm. Previous BCI research~\cite{pfurtscheller2001motor, hochberg2006neuronal} has shown that it is possible to predict limb movements (hands and feet) by analysing brain signals recorded when users imagine moving their limbs. However, for tasks like computer cursor control, this is an indirect mapping, whereas predicting the direction of intended eye movements and mapping them to control commands may be more natural. BCI systems that rely on decoding eye movements from brain activity are called `Oculomotor BCIs' or `Eye-Brain Computer Interfaces'.

In an oculomotor BCI system, the user is asked to perform voluntary eye movements while brain signals are captured over oculomotor brain regions that contribute to planning and execution of eye movements. 
For example, Jia et al.~\cite{jia2017decoding} implanted invasive microelectrode arrays in the frontal eye field and supplementary eye field of macaques, and achieved 67\% accuracy in predicting the direction of eye movements in a six-target classification task.
Tracking eye movements using brain signals also offers several advantages over camera-based eye tracking: it does not require users to remain in front of a camera, it is visually non-intrusive, it avoids issues with occlusion and background illumination, it can be integrated with virtual reality headsets~\cite{dietrich2017towards}, and it may benefit individuals with eyelid-related conditions.

Existing oculomotor studies focused on decoding `saccades', which are rapid eye movements that occur when the eyes quickly shift from one fixation point to another. Both invasive and non-invasive studies have demonstrated the capability of predicting saccade direction from brain signals. The invasive oculomotor studies implanted microelectrode arrays in macaques and showed that the direction of saccades can be predicted with high accuracies of  
66.7\%~(for 6 targets)~\cite{jia2017decoding}, 
32.2\%~(8)~\cite{graf2014brain}, 
90\%~(16)~\cite{ohmae2015decoding}, 
80.3\%~(8)~\cite{boulay2016single}, and 
43\%~(4)~\cite{lee2017decoding}. 
The brain signals were recorded from key oculomotor brain regions, such as the 
Frontal Eye Field~(FEF)~\cite{jia2017decoding, ohmae2015decoding, lee2017decoding}, 
Supplementary Eye Field~(SEF)~\cite{jia2017decoding, ohmae2015decoding, lee2017decoding}, 
Dorsolateral Prefrontal Cortex~(DLPFC)~\cite{jia2017decoding}, 
Lateral Prefrontal Cortex~(LPFC)~\cite{boulay2016single},
Lateral Intraparietal Cortex~(LIP)~\cite{graf2014brain}, 
Intraparietal Sulcus~(IPS)~\cite{lee2017decoding},
and 
Superior Parietal Lobule~(SPL)~\cite{lee2017decoding}, 
In addition to intracortical microelectrodes, Lee et al.~\cite{lee2017decoding} implanted an epidural electrocorticography (ECoG) array in macaques and reported classification accuracies of 43\%~(for 4 targets) and 80\%~(2) for saccade direction classification, respectively.
Similarly, non-invasive studies collected EEG data from able-bodied human subjects and predicted saccade directions with accuracies of 
87\%~(for 8 targets)~\cite{samadi2014eeg},  
95\%~(2) ~\cite{kastrati2021using},  and
73\%~(4)~\cite{drouin2016using}.

The experiment protocols of these studies typically implemented visually-guided, memory-guided or free-viewing saccade paradigms. In visually-guided saccades~\cite{gupta2012detecting, kastrati2021using}, the user maintains fixation at the centre of the screen and makes a saccade when a visual target appears. In delayed visually-guided saccades~\cite{ohmae2015decoding, boulay2016single, lee2017decoding}, the user continues fixating on the central dot even after the target appears, and executes the saccade only once the central dot disappears. The advantage of this delay is that it provides separate `target' and `go' cues, allowing for clearer dissociation of neural correlates related to saccade planning and execution. Another variant of this paradigm is auditory-guided saccades~\cite{funase2007single}, in which visual cues are replaced by auditory ones. In the memory-guided saccade paradigm~\cite{graf2014brain, jia2017decoding}, the user maintains fixation on a central dot while a target briefly appears and then disappears. The user must remember the target location and execute a saccade towards it only after the central fixation dot disappears. This `memory' period engages attention and working memory, increasing cognitive involvement in saccade planning. Unlike visually-guided and memory-guided saccades, which rely on external cues, free-viewing saccades~\cite{tauscher2018analysis, funase2011research} do not involve such cues; the user freely decides when and in which direction to move their eyes. The advantage of this paradigm is that it eliminates external visual or auditory influences, ensuring that the observed neural correlates are not confounded by cue-related activity.

Despite the high decoding performance, existing oculomotor studies have key limitations. Invasive studies were conducted exclusively in non-human primates, as they required electrode implantation via craniotomy, which poses clinical safety concerns for use in humans.
Consequently, there are no equivalent invasive oculomotor BCI studies in humans. In contrast, non-invasive human studies have typically relied on EOG artefacts present in EEG recordings to infer eye movement information. These artefacts originate from the corneo-retinal dipole, formed by the positively charged cornea and negatively charged retina. Eye movements cause changes in this potential, producing signals that are captured by frontal EEG electrodes. These EOG artefacts confound the underlying neural activity related to saccade planning. Existing EEG oculomotor studies~\cite{belkacem2014classification, samadi2014eeg, drouin2016using, gupta2012detecting, antoniou2021eeg, kastrati2021using} did not apply EOG correction algorithms and included data from post-saccade intervals for decoding, making it difficult to determine whether classification performance reflected true neural signals or EOG artefacts. 
These limitations highlight the need to validate the neural correlates of saccades in human subjects, without EOG artefacts, and explore their potential for oculomotor BCI applications.

The key contribution of our study is the investigation of an oculomotor BCI system in a human participant with ALS, in whom an endovascular neural interface was implanted without the need for a craniotomy. The Stentrode~\textsuperscript{TM}~\cite{oxley2016minimally} is an endovascular neural interface developed by Synchron Inc.\ (New York, USA) that records brain activity from within blood vessels. It offers signal quality comparable to epidural and subdural arrays without requiring open brain surgery~\cite{john2018signal}. The study aimed to identify the neural correlates of saccadic eye movements and decode their onset and direction using brain signals recorded via an endovascular Stentrode implanted near the supplementary motor area.
To achieve this, self-paced, Visually-Guided and Free-Viewing Saccade Tasks were designed, and the resulting data were analysed offline. Neural correlates of saccades were characterised in both the time and frequency domains.
Classification of `saccade' versus `fixation' trials (onset classification) and prediction of saccade direction (direction classification) were investigated, and the results were compared between the Visually-Guided and Free-Viewing Tasks.


%% file: 2_Methods.tex

\section{Methods}

\subsection{Participant and Implant Details}
\begin{figure}[!tbp]
    \centering
    \includegraphics[width=0.75\linewidth]{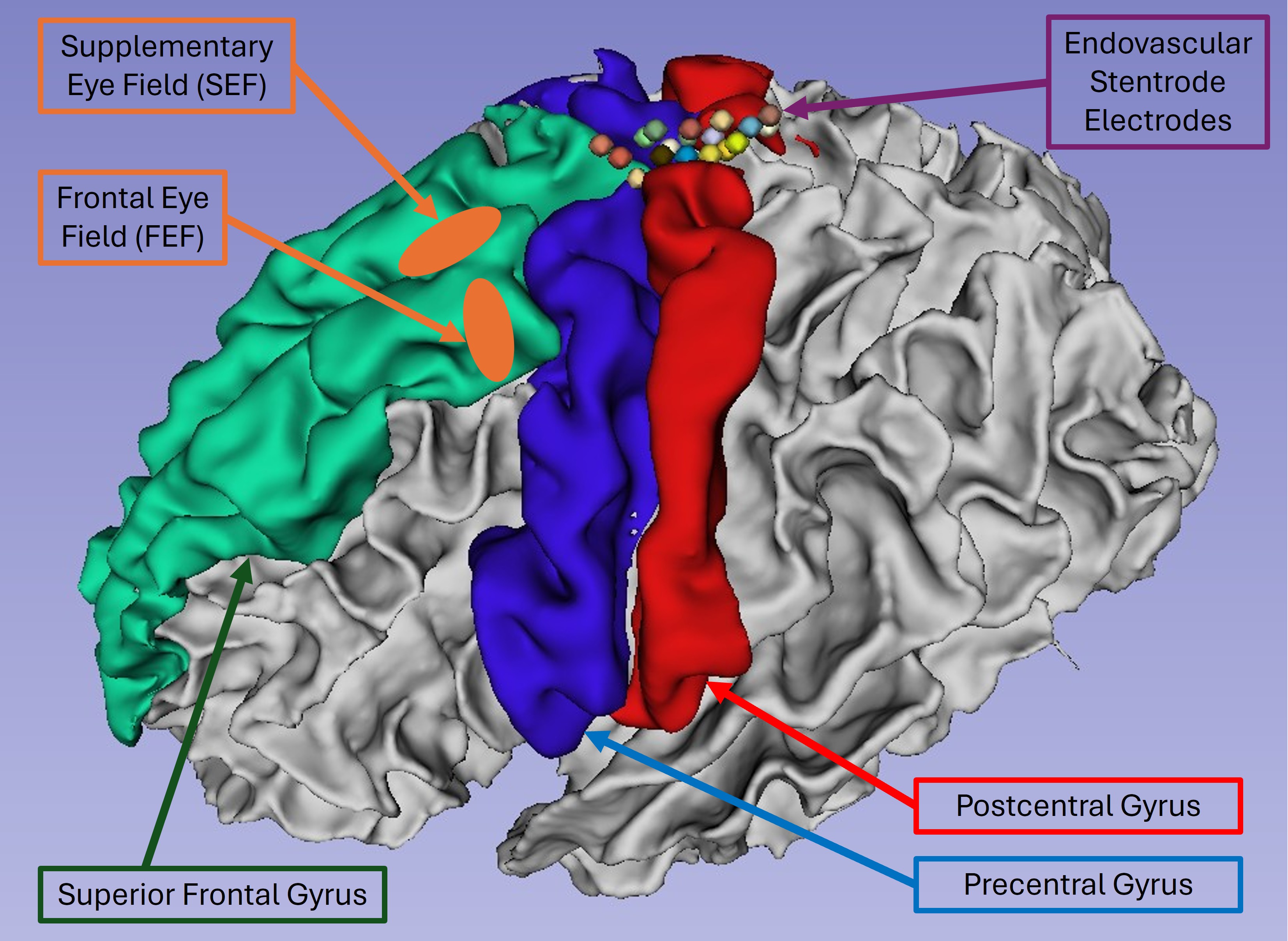}
    \caption{
    Anatomical placement of the Stentrode implant relative to major cortical landmarks in a T1-weighted MRI reconstruction of the participant’s brain.
    The electrodes are positioned near frontal oculomotor regions, including the Supplementary Eye Field~(SEF) and the Frontal Eye Field~(FEF), both of which are involved in the planning and execution of saccadic eye movements. 
    The brain anatomy and electrode positions are accurately mapped from the participant’s imaging data, whereas the SEF and FEF regions are approximate and shown for illustrative purposes based on typical anatomical landmarks.
    }
    \label{fig:electrodes-placement}
\end{figure}

Data were collected from a late-stage ALS patient (male in his 70s) with complete paralysis of the extremities. The participant was unable to speak, but retained oculomotor control and was able to use an eye tracker for computer access. The participant was implanted with an endovascular stent-electrode array~(Stentrode\textsuperscript{TM}, Synchron, NY, USA) as part of Synchron's SWITCH clinical trial~(NCT03834857), and was referred to as `P4' in an earlier study~\cite{mitchell2023assessment}. 
The Stentrode was implanted in the superior sagittal sinus, adjacent to the precentral gyrus~\cite{mitchell2023assessment}, near the Supplementary Motor Area~(SMA) of the subject.  As shown in Figure~\ref{fig:electrodes-placement}, the electrodes are anatomically proximal to frontal oculomotor regions, particularly the Supplementary Eye Field~(SEF) and the Frontal Eye Field~(FEF). 
As these regions are involved in the planning and execution of saccades~\cite{yamamoto2004human,mcdowell2008neurophysiology,pouget2015cortex}, the recorded signals may capture saccade-related neural features.

The Stentrode consisted of 16 platinum electrodes, each with a diameter of 500~\textmu m and spaced, on average, 3~mm apart, mounted on a self-expanding nitinol scaffold measuring 8~$\times$~40~mm~\cite{oxley2021motor, mitchell2023assessment, kacker2025motor}. The device was deployed through the right jugular vein and placed in the superior sagittal sinus, immediately adjacent to the precentral gyrus. The Stentrode was connected via a flexible transvascular lead to a subcutaneous Implantable Receiver-Transmitter Unit~(IRTU) located on the right side of the participant’s chest. For additional details on the implantation procedure, see~\cite{mitchell2023assessment, oxley2021motor}.


\subsection{Experiment Protocol}
\begin{figure}[!tb]
    \centering
    \includegraphics[width=0.98\linewidth]{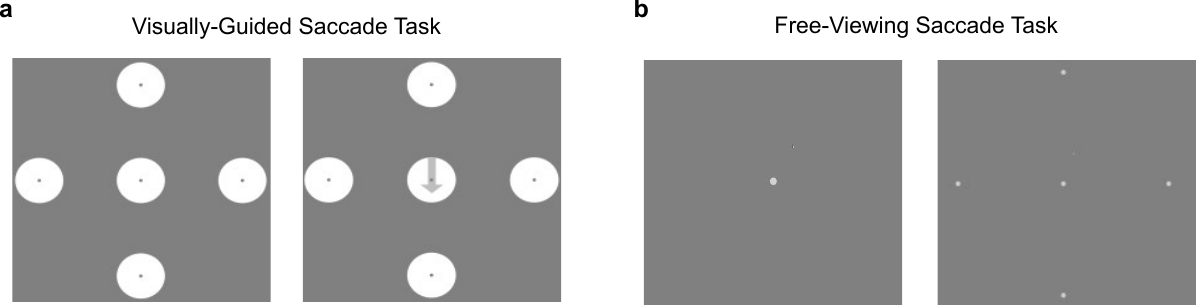}
    \caption{
    Task description
    (a)~Visually-Guided Saccade Task: Each trial began with a 1.5~s fixation display (left), followed by an arrow appearing at the centre of the screen (right) to indicate the target direction. 
    (b)~Free-Viewing Saccade Task: Fixation and saccade intervals were collected separately, with each run beginning with a 15~s fixation period (left), followed by a 2-3~min saccade interval (right) during which the participant performed self-directed saccades without external cues.
    The distance from the centre of the screen to each target location was 8.5~cm.
    }
    \label{fig:GUIs_main}
\end{figure}

Two saccade experiments were designed to study Visually-Guided and Free-Viewing saccades. This comparison was intended to reveal differences between externally cued and self-paced voluntary saccades, respectively. Both tasks were self-paced, and the user was instructed to wait briefly before performing each saccade to extend the planning interval. The user was asked to make saccades towards four different targets on the screen while data from both the eye tracker and the Stentrode were recorded simultaneously. Eye gaze information was used to label saccade onsets. Figure~\ref{fig:GUIs_main} shows the Graphical User Interface~(GUI) for both Visually-Guided and Free-Viewing Tasks.

\subsubsection{Visually-Guided Saccade Task:}
Each trial in the Visually-Guided Saccade Task~(Figure~\ref{fig:GUIs_main}(a)) began with a 1.5~s fixation interval, followed by the appearance of an arrow in the centre of the screen indicating the next gaze direction. Upon seeing the arrow, the user was instructed to wait for a short delay period before making a rapid saccade towards the target, maintain fixation there briefly, and then saccade back to the centre of the screen. Self-paced delays were introduced to minimise overlap between neural responses elicited by the visual stimulus (arrow) and subsequent eye movement. Each trial was finished when the user returned their gaze to the centre of the screen. Each trial could last up to a maximum of 10~s. If the user did not reach the target within this period, the trial was considered unsuccessful and the next trial began. For successful trials, the next trials began as soon as the user completed the bidirectional saccade.

\subsubsection{Free-Viewing Saccade Task:}
The Free-Viewing Saccade Task~((Figure~\ref{fig:GUIs_main}(b)) allowed the user to make free-viewing saccades without external cues. Each run began with a 15~s fixation interval, during which a fixation dot appeared at the centre of the screen. The user was instructed to focus on the dot without making any intentional saccades to collect data for the `fixation' trials. Blinking was allowed. Following this, a 3-4~min `saccade' interval commenced. During this phase, the user was asked to perform self-paced free-viewing saccades during which they were asked, in each trial, to look at the centre of the screen, think about the target direction, perform a saccade, briefly maintain fixation, and then return to the centre of the screen. Eye gaze data was used to identify the start and end of each trial.

\subsection{Data Collection}
Data were collected through a study approved by St. Vincent's Hospital, Melbourne~(SAGE~ID:~2022~/~PID04863-2020~/~SVH03352).
Data for the Visually-Guided Task were collected in one session and for the Free-Viewing Task in two subsequent sessions. Data recordings were made on a 15'' Surface Book~3 with a display resolution of 3240~x~2160 pixels. Neural activity was captured by the Stentrode, while eye movements were recorded using a Tobii PCEye eye tracker. The eye tracker was attached to the bottom of the Surface Book screen, with the user positioned at arm's length (approximately 50~cm) from the screen. The Tobii \href{https://us.tobiidynavox.com/products/gaze-point}{Gaze Point API} was used to control mouse movements via the eye tracker. The Stentrode data were recorded at 2000~Hz, while gaze positions were recorded at approximately 30~Hz for the Visually-Guided Task and 60~Hz for the Free-Viewing Task. The experiment was programmed in PsychoPy~\cite{peirce2019psychopy2} and the Lab Streaming Layer (LSL)~\cite{ChristianKothe} was used to synchronise the data across these three different modalities.

The eye tracker was re-calibrated at the start of each session. The cursor controlled by an eye tracker exhibited jitter, which was sometimes distracting. To mitigate this, rapid saccades were advised and the user was asked to look at the centre of the stimulus rather than the precise location of the cursor. During the Visually-Guided Task, fatigue and eye strain were reported after the session. To reduce fatigue and improve visual comfort, several modifications were made to the GUI for the Free-Viewing Task. The sizes of the target circles were reduced to facilitate fixation, and the brightness and opacity of the targets were adjusted to minimise eye strain. To further minimise fatigue, two 15~s breaks were introduced within each run. Furthermore, at the start of the Free-Viewing Task, the cursor was initially hidden; however, the user preferred to have it visible, so the cursor was made visible for the remainder of the study, as it enhanced task engagement.

For the Visually-Guided Task, a single data session was conducted with 50 trials per direction. The session included a brief 2-3~min pilot run, followed by three main runs lasting approximately 4-5~min each. Short breaks of 2-4~min were provided between runs. For the Free-Viewing Task, two data sessions were collected, with three valid runs in the first session and four valid runs in the second session. Each run consisted of 48~trials and lasted approximately 3.5-4~min. A 2-3~min break was provided between runs and the sessions were conducted approximately one week apart.

\subsection{Data Pre-processing}

\subsubsection{Labelling the Saccade Onsets}

The gaze position data recorded by the eye tracker was used to label the onset and offset of saccadic eye movements in each trial. Initially, a heuristic-based approach was used that labelled the saccade onsets based on the cursor's distance from the initial point A for each saccade between points A and B. This was followed by a visual inspection of each trial to further correct the timings of saccade onset. Trials with incomplete saccades that failed to complete a bidirectional saccade task within the maximum allowed time (10~s for the Visually-Guided and 6.5~s for the Free-Viewing Task) or those with abnormal cursor trajectories, primarily influenced by the presence of blinks, were excluded from subsequent analyses. 

Initially, 200 bidirectional trials were collected for the Visually-Guided Saccade Task, 144 trials for Session~1 of the Free-Viewing Saccade Task, and 192 trials for Session~2 of the Free-Viewing Saccade Task. 
Visually-Guided trials were the noisiest and thus had the greatest number of removed trials, with most exclusions corresponding to downward saccades. This issue arose from the angle between the eye tracker and the user's eyes causing jitter in cursor positions when the user fixated on the bottom target. Although adjusting the laptop elevation could have minimised this misalignment, we chose to maintain the natural angle that the participant typically used during their regular computer use. 
After data cleaning, 160 trials remained for the Visually-Guided Saccade Task, 130 trials for Session~1, and 170 trials for Session~2 of the Free-Viewing Saccade Task.

As part of our bidirectional saccade paradigm, we explored the differences between the initial saccade toward a target and the second saccade when the user looked back to the original fixation point. The former is called the `initial saccade' and the latter is called the `back saccade'. Previous oculomotor BCI studies have typically focused on the initial saccade, but our aim was to investigate the differences in features related to both types of saccades. We considered three cases: (i)~initial saccades separately, (ii)~back saccades separately, and (iii)~combined saccades, treating both initial and back saccades as a single event. Combining saccades effectively doubled the number of unidirectional saccade trials.

To clarify the labelling scheme, consider a trial in which the user initially makes a saccade toward a target appearing near the top corner of the screen. This would be labelled as `up initial saccade'; when the user looks back to the centre of the screen to complete the trial, it would be labelled as `down back saccade'. Similarly, a `left initial saccade' would have a complementary `right back saccade. In the combined saccades case, a `up combined saccade' includes both the `up initial saccade' and the `up back saccade'.

\subsubsection{Extracting Fixation and Saccade Trials}
In the case of the Visually-Guided Task, each saccade trial was preceded by a 1.5~s fixation trial. However, no dedicated pre-saccade fixation interval was present for free-viewing saccades. Therefore, fixation epochs were extracted during the 15~s fixation period of each run of the Free-Viewing Task. The Visually-Guided Task, by default, had an equal number of fixation and saccade segments. In order to facilitate comparison, for the Free-Viewing Task, a moving window with a specific stride size was employed, ensuring that the total number of fixation trials matched the number of saccade trials. In some cases, the adjacent fixation trials of the Free-Viewing Task overlapped with each other. However, block-wise cross-validation was applied to ensure that the data from each run were used exclusively for training or testing, thus minimising the bias of temporal correlations between overlapping fixation trials~\cite{white2023k}. Additionally, while eye blinks may have occurred during fixation trials, particularly in the Free-Viewing Task, all fixation trials were included in the analysis due to the limited data available, regardless of blink presence.

\subsubsection{Sensor Space vs.\ Source Space Data:}
As neural activity across different channels of the original Stentrode recordings was highly correlated, Independent Component Analysis (ICA)~\cite{makeig1995independent} was applied to separate neural signals arising from distinct neural sources. This also helped to effectively separate cardiac artefacts from the neural signals. ICA is a blind source separation method that decomposes sensor-level data into statistically independent components, which we refer to as the source space. The original electrode recordings are referred to as the sensor space. ICA assumes the observed brain signals $\boldsymbol{X}(t) \in \mathbb{R}^{n \times T}$, where $n$ is the number of channels and $T$ is the number of time points, are linear mixtures of statistically independent sources $\boldsymbol{S}(t)$, combined through an unknown mixing matrix $\boldsymbol{A}$, such that $\boldsymbol{X}(t) = \boldsymbol{A} \cdot \boldsymbol{S}(t)$. 
An unmixing matrix $\boldsymbol{W} \approx \boldsymbol{A}^{-1}$ is estimated and applied to the sensor space data to obtain the independent components in source space: $\hat{\boldsymbol{S}}(t) = \boldsymbol{W} \cdot \boldsymbol{X}(t)$. 
In this study, ICA was implemented using the MNE-Python library (version 1.2.1)~\cite{gramfort2013meg}, employing the FastICA algorithm~\cite{ablin2018faster}, which estimates $\boldsymbol{W}$ by maximising non-gaussianity of the recovered components under unit-variance constraints.
Before applying ICA, the data were visually inspected, and two noisy channels were removed from the 15 active channels. The remaining 13 channels were then used to estimate the Independent Components (ICs). A 1~Hz high-pass zero-phase FIR filter was applied to preprocess the data before computing the ICA unmixing matrix.



For the Free-Viewing Task, two data sessions were recorded, and the ICs calculated for each of these sessions may not maintain a one-to-one correspondence. To align the corresponding ICs across sessions, a group-level ICA approach~\cite{huster2015group, eichele2011eegift} was used, in which data from both sessions were concatenated to compute a single ICA unmixing matrix. This matrix was then applied separately to each session. Alternative approaches, such as visualising power spectra or topographical maps of individual ICs or applying clustering techniques~\cite{grandchamp2012stability}, have been used in literature to establish IC correspondence in EEG data. However, because the topographical maps derived from the Stentrode are not yet well characterised in the literature, such an investigation was beyond the scope of the present study.

\subsection{Signal Processing Pipeline}
\label{ssec:SP-Pipeline}
\begin{figure}[!tb]
    \centering
    \includegraphics[width=1\linewidth]{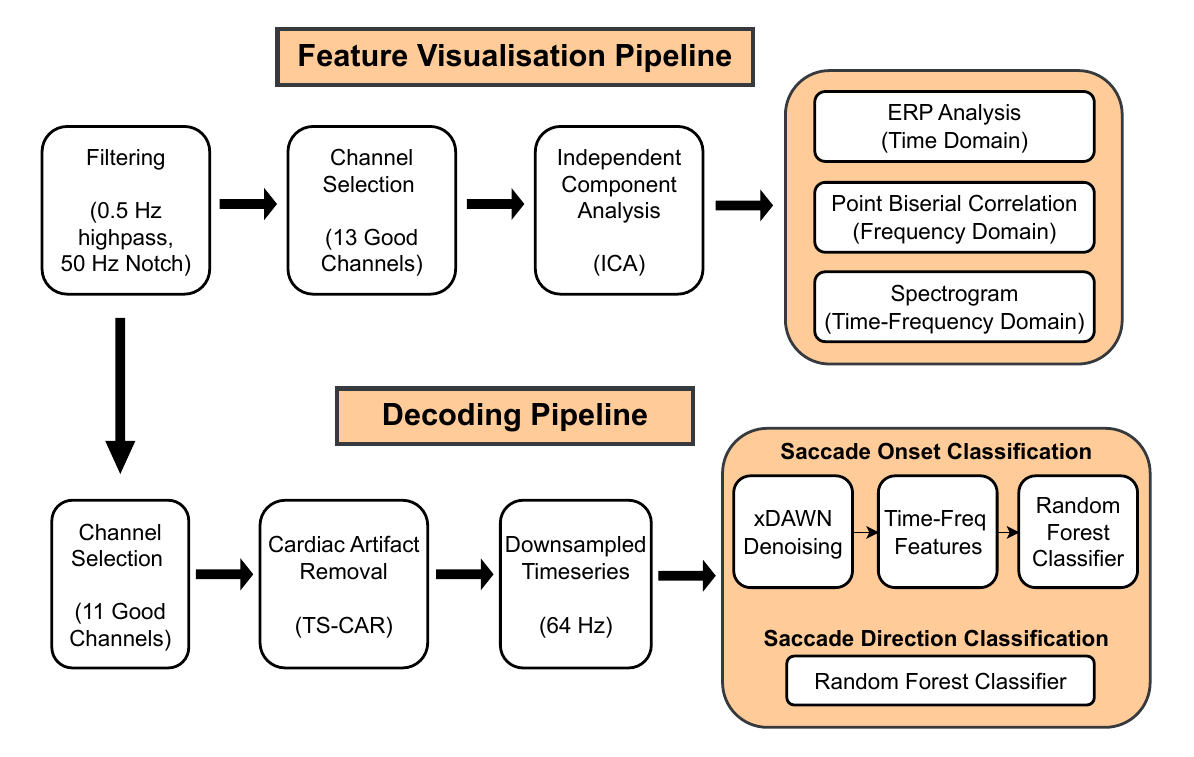}
    \caption{Signal processing pipelines for feature visualisation and decoding.
    (Top)~The Feature Visualisation Pipeline visualised neural features in source space, involving filtering, channel selection, Independent Component Analysis (ICA), and analysis in the time, frequency, and time-frequency domains.
    (Bottom)~The Decoding Pipeline performed classification in sensor space, involving filtering, channel selection, cardiac artefact removal, and downsampling. For saccade onset classification, xDAWN denoising and time-frequency feature extraction were employed before classification with a random forest classifier, whereas saccade direction classification achieved better performance by directly classifying downsampled data using a random forest classifier.}    
    \label{fig:Pipeline}
\end{figure}

Figure~\ref{fig:Pipeline} illustrates the signal processing pipeline for both feature visualisation and classification. A key difference is that feature visualisation was performed in the source space, whereas classification focused mainly on sensor space data. The use of sensor space data for decoding was chosen to facilitate real-time processing and mitigate the instability of ICs in real-time applications, as ICs computed across different sessions may not have direct one-to-one correspondence. 

The first step, filtering the data, was common to both pipelines: a  0.5~Hz highpass filter was applied to remove low-frequency drifts and DC offset, and a 50~Hz notch filter was applied to suppress line noise from the Stentrode recordings. Both filters were implemented as zero-phase FIR filters using a Hamming window with forward-backward filtering. The filter order was automatically determined according to MNE-Python's default `firwin' method~\cite{gramfort2013meg}.
The following subsections discuss the later steps in each pipeline separately.

\subsubsection{Feature Visualisation}
\label{sssec:FV}

The temporal, spectral, and time-frequency features were visualised using event-related potentials, point-biserial correlation coefficients ($r^2$), and spectrograms, respectively. A brief description of each technique is provided below.

\paragraph{Evoked Responses (ERP Analysis):}
Evoked responses, or event-related potentials (ERPs), represent the average neural activity across multiple trials, time-locked to a specific event, such as the onset of a cue or saccade. Since trial averaging primarily preserves low-frequency components, a 15~Hz low-pass zero-phase FIR filter was applied to smooth the ERP plots for visualisation. In addition, individual trials were plotted to observe ERP patterns at the single-trial level.

\paragraph{Biserial Correlation Coefficient (\( r^2 \)):}
The squared point biserial correlation coefficient (\( r^2 \)) \cite{sannelli2019large} was used to identify discriminative frequency bands for the fixation vs.\ saccade trials. It is a correlation between a real variable (a feature such as power spectral density) and a dichotomous variable containing class information. This metric quantifies the variance of the signal related to its class affiliation. The coefficient is given by
\begin{equation}
r(x_1, x_2) = \frac{\sqrt{N_1 N_2}}{N_1 + N_2} \frac{\mathrm{mean}(x_1) - \mathrm{mean}(x_2)}{\mathrm{std}(x_1 \cup x_2)},
\end{equation}
where \( x_1 \) and \( x_2 \) represent the power spectral density values in a given frequency bin for each condition and \( N_1 \) and \( N_2 \) denote the numbers of trials. The value of \( r \) can be positive or negative. Since we are only interested in the magnitude of the difference, regardless of direction, it is common practice to report the squared values (\( r^2 \)) to ensure that all values are positive. An \( r^2 \) value of 1 indicates that the feature is perfectly correlated with the class, while an \( r^2 \) value close to 0 suggests that the feature is not discriminative for the two classes.

In our analysis, point-biserial correlation coefficients ($r^2$) were computed between the power spectral densities of fixation and saccade trials. The power spectral densities were estimated using the multitaper method with a frequency resolution of 1~Hz. The $r^2$ score was then calculated for each frequency bin to quantify the discriminability between the two classes across different frequency bands. 
Based on exploratory analysis, saccade-related features were predominantly observed in the low-frequency range (0–30~Hz); therefore, $r^2$ analysis was restricted to this frequency band.


\paragraph{Spectrograms:}
Spectrograms show the spectral power as it evolves over time and provide a useful means of visualising dynamic spectrotemporal features. Data were downsampled to 1024~Hz, and time-frequency power estimates were calculated in the 1-30~Hz range using 0.1~Hz frequency bins. Time-frequency decomposition was performed using Morlet wavelets, and the mean power across trials was computed for each condition. To express power values on a decibel~(dB) scale, baseline normalisation was applied. Specifically, each power value was divided by the mean baseline power, followed by a logarithmic transformation and multiplication by 10. The baseline interval was defined as the full duration of the trial, extended by an additional 5~seconds preceding the trial onset.

\subsubsection{Saccade Classification}
Decoding saccadic eye movements can be categorised into two categories.
    `Saccade Onset Classification' is a binary classification problem, and the goal is to distinguish the brain activity corresponding to saccade trials from fixation trials. 
    This signal processing approach is also called `fixation vs.\  saccade' classification. 
    `Saccade Direction Classification' predicts the target location (i.e., the intended saccade direction). In this study, there were four different targets: left, right, up, and down directions. In addition to four-class decoding results, we also report binary classification results (left vs.\ right, up vs.\ down, etc.) as previous studies~\cite{jia2017decoding, nategh2020decoding} have shown that oculomotor BCIs can be more selective to targets in certain directions.

Before applying decoding algorithms on sensor space data, a Time-Selective Common Average Referencing (TS-CAR) algorithm was applied to remove cardiac artefacts. Unlike standard common average referencing, which applies correction across all time points, TS-CAR selectively targets time points containing cardiac artefacts, minimising its impact on non-artefact regions~\cite{bailey2024eeg}. Cardiac artefacts in Stentrode recordings appear as large, temporally aligned peaks across channels~\cite{Bennett2021}, making them suitable for suppression through spatial averaging. However, for TS-CAR to be effective, artefacts must exhibit comparable amplitudes across channels. Following visual inspection, two of the 13 initially selected channels were excluded due to lower artefact and neural signal amplitudes compared to other channels. TS-CAR was applied within 130~ms windows around each detected R-peak, leaving other signal segments unaffected.  Additionally, to preserve the rank of the channels, we first added an extra zero channel, applied common average referencing, and then removed the extra channel. This step ensured that the overall rank of the data was maintained, avoiding errors in decoding algorithms that rely on the full rank of the data~\cite{MakotoPreprocessing}.

To extract features, the data were downsampled to 64~Hz, as it gave results comparable to those of decoding data at the original sampling rate (2000~Hz). Then, for the `saccade onset classification', the xDAWN denoising algorithm~\cite{rivet2009xdawn, rivet2011theoretical} was applied to increase the SNR of ERP features, followed by extracting features in both time and frequency domains. xDAWN is a spatial filtering method to increase the Signal-to-Signal-plus-Noise Ratio~(SSNR) of the evoked responses. It maps the original sensor space data to source space and uses a supervised approach to select the best xDAWN components that will maximise the contrast between target trials and non-target trials. The application of xDAWN denoising effectively reduces the number of channels while increasing the SSNR of the evoked responses, which simplifies the classification. In contrast, for `saccade direction classification', it was found that not applying xDAWN and decoding the original time-series data gave slightly better performance.

To extract time-domain features, each trial was divided into non-overlapping 100~ms segments and the mean, standard deviation, variance, kurtosis, and root-mean-square (RMS) values were computed for each segment. For frequency domain feature extraction, the mean and standard deviation of spectral power were calculated in the delta (0.5-4~Hz), theta (4-8~Hz), alpha (8-12~Hz), and beta (12-30~Hz) bands using the multitaper method.

For classification, multiple algorithms were evaluated, with the Random Forest classifier performing slightly better than the others (see~\ref{Appendix-Comparison-Classifiers} for details). 
To account for class imbalances resulting from the removal of bad trials, the results are reported using AUC-ROC scores rather than accuracy.
The decoding algorithms were mainly implemented using the MNE-Python~\cite{gramfort2013meg} and Scikit-learn~\cite{scikit-learn} libraries in Python. The xDAWN denoising algorithm accepted the `n\_components' argument, which yielded the best decoding results when set to 2 components. The output of xDAWN was standardised to the range~[0, 1] using min-max scaling, after which the scaled features were fed into a random forest classifier for classification. Grid search was performed to optimise the `n\_estimators' and `max\_depth' hyperparameters of the random forest classifier.

\paragraph{Performance Evaluation:} 
The performance of a BCI decoder can be evaluated using either a within-session or cross-session decoding setup. Since the Visually-Guided Task included only one session, we reported only within-session results for this task, while both within-session and cross-session results were reported for the Free-Viewing Task.

A Leave-One-Run-Out Cross-Validation (LORO CV) approach, which is a special case of block-wise CV, was used for both within-session training and testing, as well as cross-session training. In block-wise CV, data is divided into $p$ non-overlapping blocks and, in each split, trials from one block are held as a test set while trials from the rest of the blocks are used as the training set. This procedure is repeated $p$ times until all blocks appeared as the test set exactly once. Leave-One-Run-Out is a specific case of block-wise CV where each block corresponds to each run (small consecutive data collection sub-sessions within each session). The benefit of this approach is that all trials within a block remained together in either the training or the testing set, and thus temporal correlations among temporally nearby samples would not confound the results~\cite{west2023machine}. This contrasts with a classical K-Fold CV, where trials are randomly assigned to training and test splits, potentially overestimating the results due to the high correlation among samples caused by their proximity in time~\cite{white2023k}.

%% file: 3_Results.tex





\section{Results}
\label{sec:results}

\subsection{Time Domain Features: ERP Analysis}
\label{ssec:time-domain-features}

\begin{figure}[!tbp]
    \centering
    \includegraphics[width=1\linewidth]{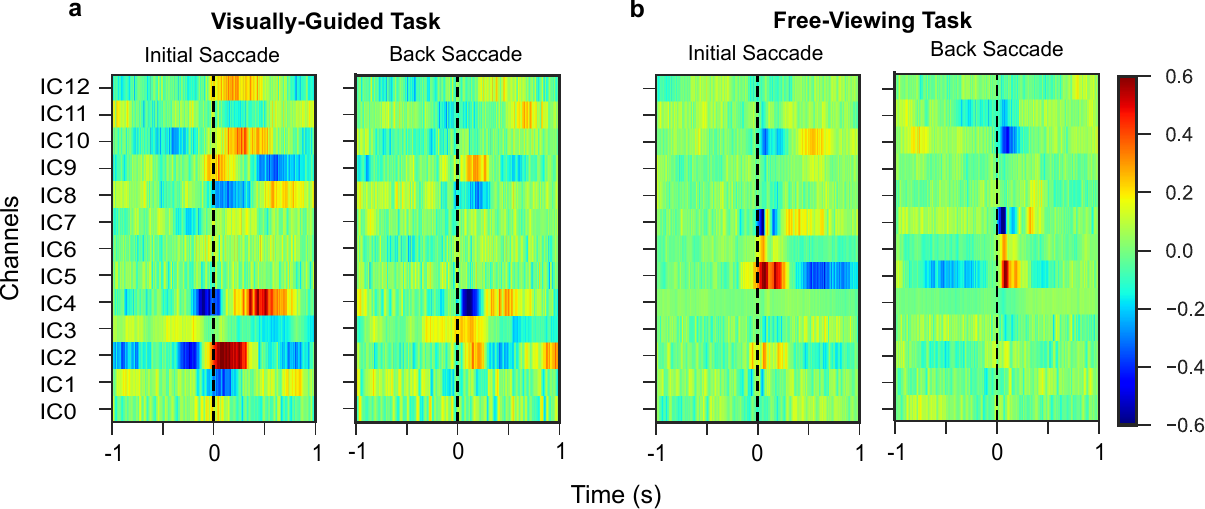}
    \caption{Evoked responses of Independent Components~(ICs) plotted against time, aligned to saccade onsets (vertical black lines at $t=0$) for both initial and back saccades. 
    Red and blue represent positive and negative deflections, respectively.
    (a)~Visually-Guided Task (160 Trials): IC2 and IC4 showed positive and negative deflections, respectively, around the initial saccade, while IC4 also showed negative deflection following the back saccade. 
    (b)~Free-Viewing Task (303 trials): IC5 showed positive deflection, whereas IC7 and IC10 showed negative deflections around both initial and back saccades.}
    \label{fig:ic_selection_ERPs}
\end{figure}

To visualise the discriminative channels in source space, the plots in Figure~\ref{fig:ic_selection_ERPs} show evoked responses of Independent Components~(ICs) around saccade onsets~($t=0$) for both Visually-Guided and Free-Viewing Tasks. The discriminative ICs exhibited positive or negative deflections, corresponding to increases or decreases in amplitude, respectively, and are shown in red and blue around the onset of the saccade ($t=0$).

For the Visually-Guided Task~(Figure~\ref{fig:ic_selection_ERPs}(a)), IC2 and IC4 showed distinct ERPs around saccade onsets. For the initial saccade, the evoked response of IC2 showed a change from negative amplitude (blue) to positive amplitude (red) around $t=0$, while IC4 showed a negative ERP component around $t=0$. For the back saccade, IC4 showed a negative deflection after the onset of the saccade. It was also observed that the evoked responses of the initial saccade generally had stronger ERP components in different channels compared to the back saccade, which may suggest a stronger signal for the initial saccade.

For the Free-Viewing Task~(Figure~\ref{fig:ic_selection_ERPs}(b)), channels IC5, IC7 and IC10 showed distinct activity around saccade onsets. The ERPs for both initial and back saccades were very consistent, IC5 showing a positive ERP component, while IC7 and IC10 showed a negative ERP component around $t=0$. When comparing the initial and back saccades, the positive ERP deflection for the initial saccade lasted longer than the back saccade, while IC10 in the back saccade showed stronger ERP components represented by a darker shade of blue around $t=0$.

\begin{figure}[!htbp]
    \centering
    \includegraphics[width=0.94\linewidth]{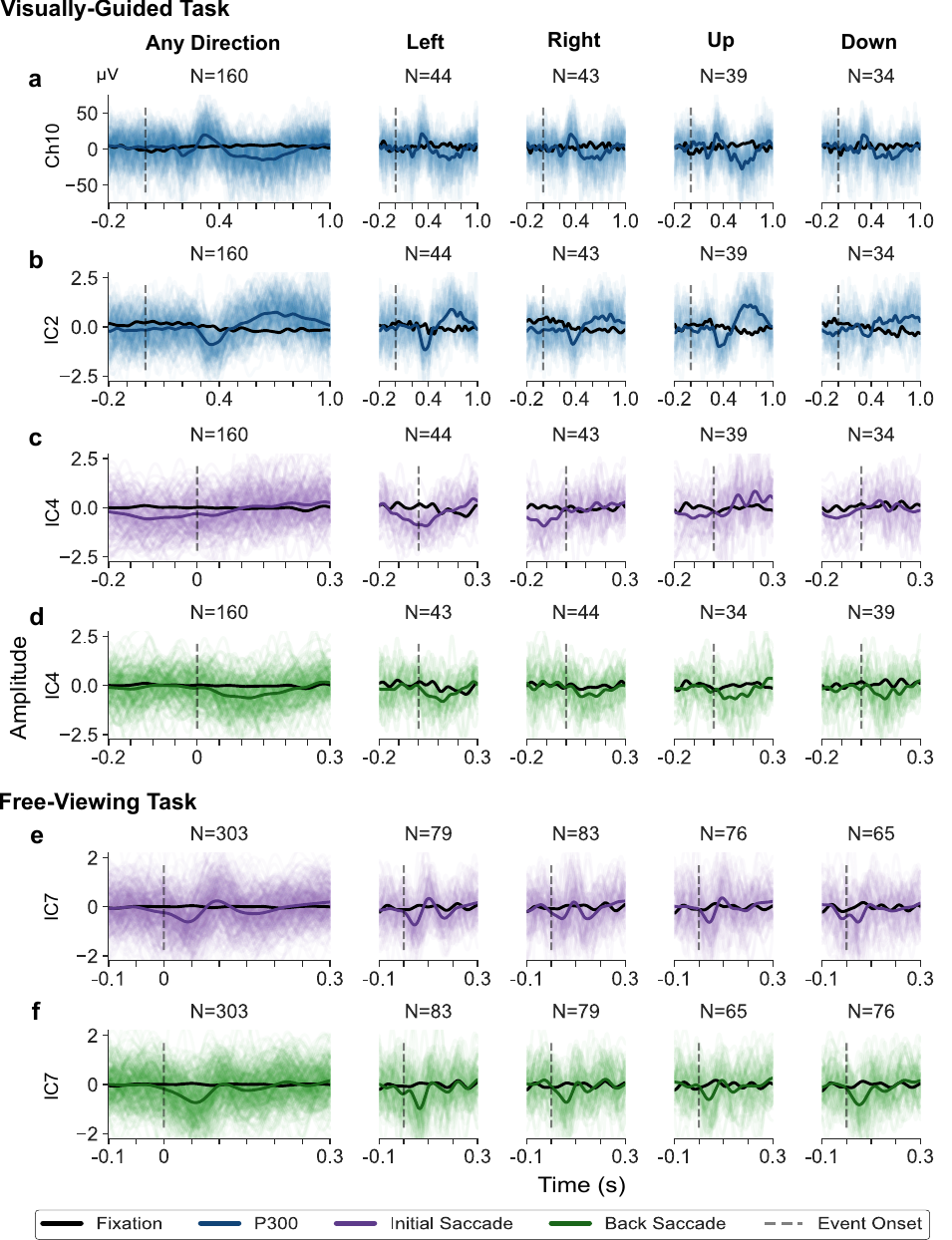}
    \caption{Evoked responses for the Visually-Guided (a-d) and Free-Viewing Tasks (e-f), shown across directions (columns) and different events (rows).  
    (a, b)~P300 potentials time-locked to cue onset ($t=0$) using sensor space (Channel~10) and source space (IC2) data.  
    (c-f)~Saccade-related ERPs time-locked to initial (c, e) and back (d, f) saccades using IC4 (Visually-Guided) and IC7 (Free-Viewing). 
    Thin lines show individual trials, while bold lines show ERP averages across `N' trials. Blue, purple, green, and black lines represent P300, initial saccade, back saccade, and fixation trials, respectively.}
    \label{fig:ERP_Main}
\end{figure}

The next step was to visualise the ERPs in different directions for both tasks using the most discriminative channels. Figure~\ref{fig:ERP_Main} shows the evoked responses of different events (P300, initial saccade, back saccade), for direction-specific (left, right, up, down) and direction-agnostic (any direction) trials. The `any direction' column represents ERPs of data pooled across all directions. Panel (a) represents the ERP of channel 10 in the sensor space, while the remaining panels (b)-(f) correspond to the source space data. In each ERP subplot, the amplitude is plotted against time, with a grey vertical line at $t=0$ representing event onset (arrow, initial saccade, back saccade). `N' represents the number of trials across which the ERP average was computed. Furthermore, the black bold line represents the ERP of the fixation trials, which serve as a baseline for comparison, whereas the bold blue, purple, and green lines represent ERPs of P300, initial saccade, and back saccade trials, with thin traces showing individual trials. 
It is worth noting that the ERPs in different panels are plotted on different timescales, chosen to ensure that the ERP components return to baseline while remaining short enough to allow visualisation of features in individual trials.

\subsubsection{P300 Signal in Visually-Guided Task:}
\label{sssec:p300}
Figure~\ref{fig:ERP_Main}(a) and Figure~\ref{fig:ERP_Main}(b) show evoked responses time-locked to the appearance of visual cue (arrow) for sensor space and source space data, respectively. For the ERP of channel~10 (Ch10) in the sensor space (Figure~\ref{fig:ERP_Main}(a)), there exists a strong positive peak (bold blue line) approximately 300 to 400~ms after arrow onset, which is characteristic of a P300 response to visual stimuli~\cite{picton1992p300}. In contrast, the ERP of the fixation trials (bold black line) did not show a distinct ERP pattern. The individual trial traces (thin blue lines) further demonstrated that the P300 response was consistently present at individual trials, indicating a high signal-to-noise ratio~(SNR) for these potentials. It is important to note that while P300 was plotted for channel~10, due to the high inter-channel correlation in sensor space data, similar P300 potentials were observed over multiple channels.

Figure~\ref{fig:ERP_Main}(b) shows the P300 potential in the source space using IC2. The timings of these cue-evoked potentials matched those of sensor space potentials (Figure~\ref{fig:ERP_Main}(a)) with the only difference being that the polarity of the response was flipped, which is a known property of the ICA algorithm since it seeks statistically independent sources, and the mathematical formulation is indifferent to the sign of the components~\cite{miyakoshi2024one}.

Although the ERPs of the data pooled in all directions (`any direction') showed prominent P300 potentials, similar potentials were also observed in four individual directions (left, right, up, and down) with P300 peaks occurring in 300-400~ms interval following cue onset, confirming that the response was consistently elicited by the visual stimulus and was independent of the intended saccade direction. However, the ERPs for individual directions were comparatively noisier, and in some cases the ERP of the fixation trials (bold black line) was not completely flat. This could be attributed to the lower SNR, as averaging fewer trials per direction reduces the effectiveness of noise suppression.

Another important observation was that, in each visually guided trial, the participant waited briefly after the arrow appeared and then performed the saccade. As a result, saccade-related potentials appear after the cue-evoked P300 response. The ERP peaks observed in the 400-800~ms range following the P300 likely reflect this saccade-related activity.

\subsubsection{Saccadic Potentials in Visually-Guided Task:}

Figure~\ref{fig:ERP_Main}(c) and Figure~\ref{fig:ERP_Main}(d) show saccade-related potentials for the initial and back saccades, respectively, for the Visually-Guided Task. For the initial saccade (Figure~\ref{fig:ERP_Main}(c)), the ERPs (purple) appeared different in different directions, likely due to overlap with previous P300 potentials due to the visual cue. However, for the back saccade (Figure~\ref{fig:ERP_Main}(d)), there appeared somewhat consistent ERP components that began around the onset of the saccade and persisted for approximately 200-300~ms after saccade execution. Furthermore, even for the back saccade, the individual trials are more noisy and therefore had a lower SNR compared to the P300 responses observed in Figure~\ref{fig:ERP_Main}(a-b).

\subsubsection{Saccadic Potentials in Free-Viewing Task:}

Figure~\ref{fig:ERP_Main}(e) and Figure~\ref{fig:ERP_Main}(f) show saccade-related potentials of initial and back saccades, respectively, for the Free-Viewing Task. The ERPs of saccade trials appeared very similar for both types of saccade. This was likely because the Free-Viewing Task did not involve any external cues and the self-paced nature of the task allowed both initial and back saccades to evoke similar neural responses. As observed in ERPs of pooled data (`any direction'), the ERPs (bold purple green lines) began within approximately 50~ms before saccade onset and persisted for up to 200~ms after saccade execution, with the onset marked by the grey vertical dashed line at $t=0$. The ERP of the fixation trials (bold black line) remained flat, and individual saccade trials (thin purple and green traces) exhibited highly consistent patterns, indicating a high SNR for the saccadic potentials in the Free-Viewing Task. Furthermore, similar ERP components were observed across different saccade directions, with negative ERP peaks observed just after saccade execution. However, the amplitudes and latencies of these potentials varied slightly between directions, which may indicate the presence of direction-specific neural responses.

Although only ERPs of the most prominent channels are discussed here, Figure~\ref{fig:ERP_Appendix} in \ref{Appendix-Saccadic-Potentials} shows ERPs of other good channels in the source space. For the Visually-Guided Task, the ERPs of IC4 (Figure~\ref{fig:ERP_Appendix}(a)) demonstrated  cue-related potentials occurring approximately 400-650~ms after arrow appearance. Furthermore, the ERP of IC2 for the initial saccade~(Figure~\ref{fig:ERP_Appendix}(b)) lacked a consistent ERP component, whereas for the back saccade~(Figure~\ref{fig:ERP_Appendix}(c)), the ERP of IC2 appeared as a positive peak that began just after the start of the saccade and lasted for about 250-300~ms after back saccade onset.  For the Free-Viewing task, IC5 showed a positive ERP for both initial and back saccades, which began just before the onset of the saccade for the initial saccade (Figure~\ref{fig:ERP_Appendix}(d)) and on saccade onset for the back saccade (Figure~\ref{fig:ERP_Appendix}(e)) and lasted for approximately 200-300~ms after saccade onset. ERPs in different directions and single-trial plots showed somewhat similar patterns for free-viewing saccades.

\subsection{Frequency Domain Features: Biserial Correlation Coefficient}
\label{ssec:freq-domain-features}

\begin{figure}[!htbp]
    \centering
    \includegraphics[width=1\linewidth]{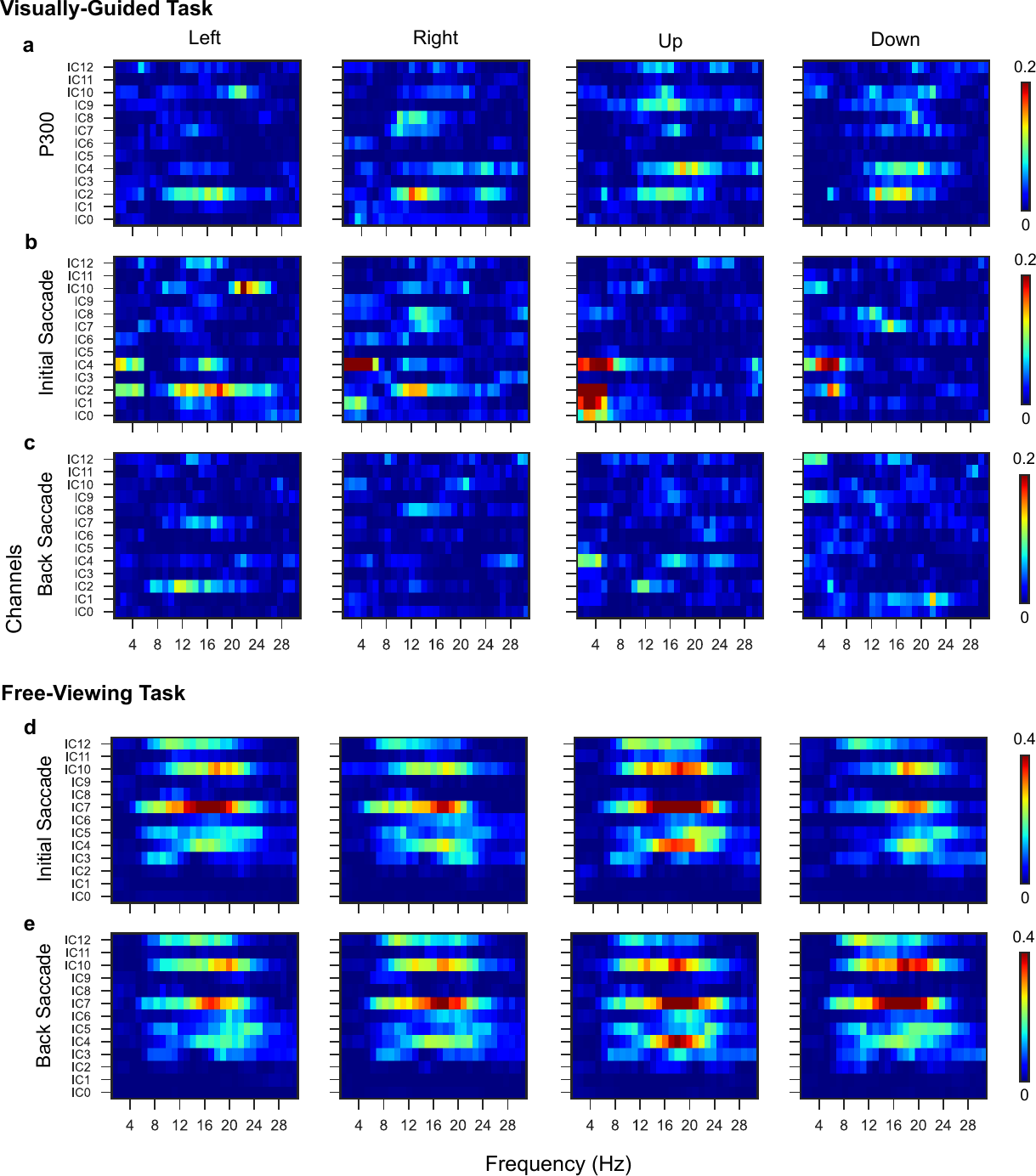}
    \caption{Squared Point Biserial Correlation Coefficient ($r^2$) plots between power spectral densities of different events against fixation trials for both Visually-Guided and Free-Viewing Saccade Tasks. Rows correspond to different events (P300, initial saccade, back saccade), while columns correspond to saccades trials in left, right, up and down directions. Each subplot shows $r^2$ of different channels against different frequencies. Higher $r^2$ values are shown in red and represent frequencies where the events of interest show disriminable activity when compared with fixation trials. IC2 and IC4 for the Visually-Guided Task, and IC4, IC7, IC10 and IC12 for the Free-viewing Task showed high $r^2$ scores.}
    \label{fig:R2_individual_direction}
\end{figure}

The point biserial correlation coefficients ($r^2$) were calculated to identify which frequency bands provided the most discriminative features between the power spectral densities of the fixation trials versus the saccade (or P300) trials. Figure~\ref{fig:R2_individual_direction} shows the $r^2$ scores plotted as heat maps for each channel (source space) plotted against frequencies for different directions (columns: left, right, up, down) and different events (rows: P300, initial saccade and back saccade) for both Visually-Guided and Free-Viewing Tasks. Warmer colours (yellow to red) in the plots indicate higher $r^2$ values, reflecting higher discriminative power, while cooler colours (blue) represent lower $r^2$ values, corresponding to features with limited class separability.

For the Visually-Guided Task, Figures~\ref{fig:R2_individual_direction}(a),~\ref{fig:R2_individual_direction}(b), and~\ref{fig:R2_individual_direction}(c) show the $r^2$ scores for all channels (y-axis) for the conditions(arrow onset, initial saccade and back saccade, respectively), with IC2 and IC4 showing high $r^2$ scores (red coloured) compared to other channels. 
For the P300 trials (Figure~\ref{fig:R2_individual_direction}(a)), the discriminative features appeared in IC2 and IC4, primarily in the 8-24~Hz range, with IC2 showing consistent scores across all four directions. In contrast, the initial saccades (Figure~\ref{fig:R2_individual_direction}(b)) exhibited discriminative features in the lower frequency range (below 8~Hz), with IC4 being more consistent across directions. The back saccades (Figure~\ref{fig:R2_individual_direction}(c)) showed relatively small $r^2$ scores compared to the other two conditions. As the disriminative frequency bands were different when comparing P300 versus fixation, against saccade versus fixation, this suggests the presence of unique spectral features for P300 and visually-guided saccades.

For the Free-Viewing Task, Figures~\ref{fig:R2_individual_direction}(d) and~\ref{fig:R2_individual_direction}(e) present the $r^2$ scores for the initial and back saccades, respectively. Identical $r^2$ patterns across different channels were observed for all directions and both types of saccades. IC7 gave the highest $r^2$ scores, followed by IC10, IC4, and IC12. The most discriminative frequencies were found within the 6–25~Hz range, with peak discrimination occurring between 12–24~Hz. A general observation was that free-viewing saccades provided higher $r^2$ scores compared to visually-guided saccades, as reflected in the scale limits: $r^2$~=~0.4 for the Free-Viewing Task and $r^2$~=~0.1 for the Visually-Guided Task. This difference may be attributed to the experimental structure: unlike the Visually-Guided Task, where each saccade trial was preceded by a fixation trial, all fixation trials in the Free-Viewing Task occurred together, followed by all saccade trials. This temporal separation might have made the discrimination between fixation and saccade trials easier in the Free-Viewing Task.

\subsection{Time-Frequency Domain Features: Spectrograms}
\label{ssec:spectrogrograms}

\begin{figure}[!htbp]
    \centering
    \includegraphics[width=0.95\linewidth]{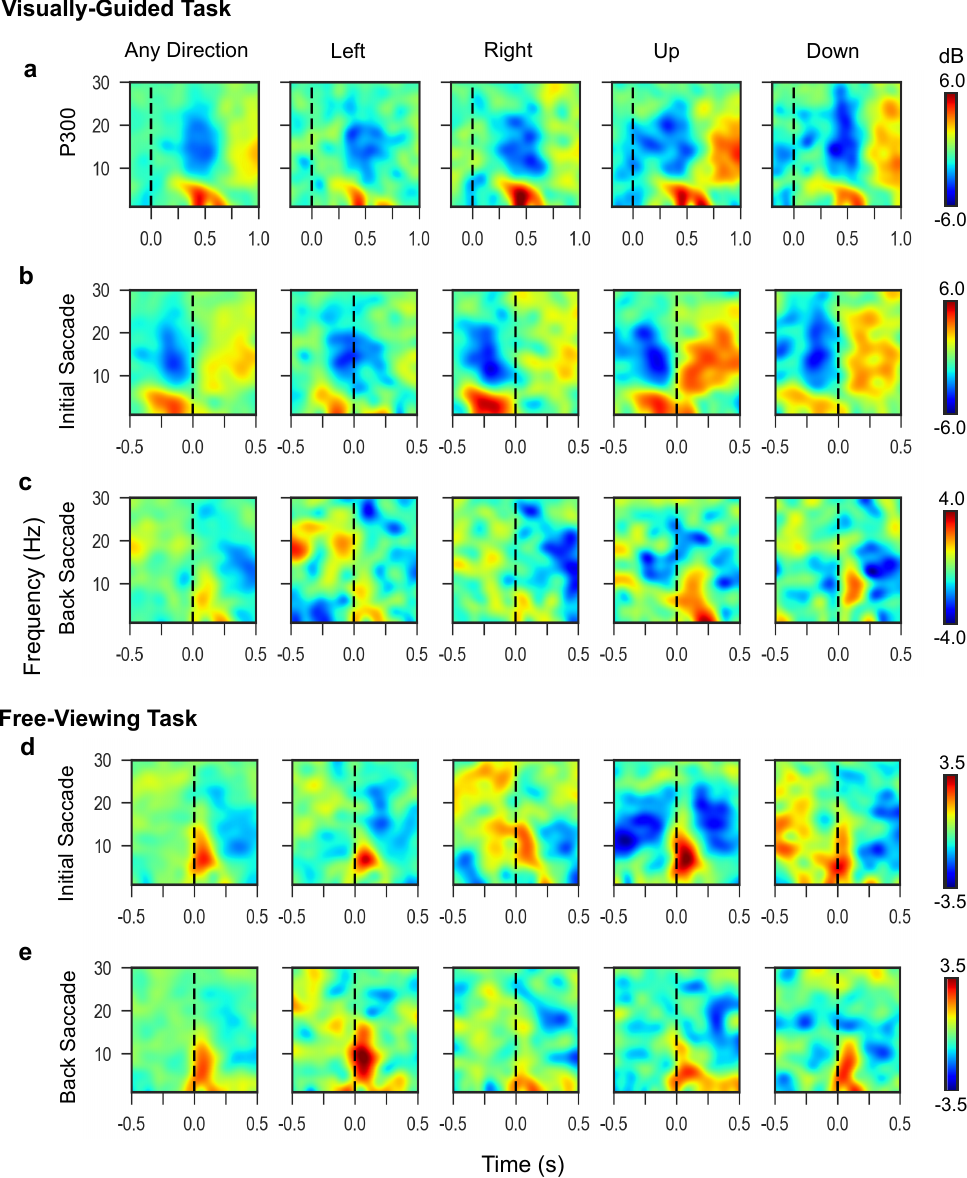}
    \caption{Spectrograms of Visually-Guided Task (IC4; panels~a–c) and Free-Viewing Task (IC7; panels~d–e), time-locked to key events (rows) across different saccade directions (columns). The black vertical line at $t=0$ marks event onset (arrow cue, initial saccade, or back saccade). In the Visually-Guided Task, low-frequency synchronisation (red) and beta desynchronisation (blue) were observed following the cue~(a) and before initial saccade~(b). In the Free-Viewing Task, prominent low-frequency synchronisation (below 15~Hz) was observed around both initial and back saccades~(d–e).}
    \label{fig:spectrograms_main}
\end{figure}

Figure~\ref{fig:spectrograms_main} presents the spectrograms of IC4 and IC7 from the Visually-Guided and Free-Viewing Tasks, respectively. 
Each row corresponds to the spectrogram of a specific independent component, time-locked to a relevant event of interest (arrow onset, initial saccade, or back saccade) at $t = 0$. For each event type, spectrograms are shown separately for the four saccade directions (left, right, up, and down), along with a combined spectrogram (any direction) to highlight features that were consistent across directions. 
In the spectrograms, red and blue indicate increases and decreases in spectral power (in decibels), respectively, relative to the baseline period. These changes are referred to as synchronisation and desynchronisation, respectively.

For the Visually-Guided Saccade Task, Figures~\ref{fig:spectrograms_main}(a), \ref{fig:spectrograms_main}(b), and \ref{fig:spectrograms_main}(c) show the spectrograms for IC4, time-locked to the appearance of the visual cue, the initial saccade, and the back saccade, respectively. 
In Figure~\ref{fig:spectrograms_main}(a), the P300 potentials appeared as a synchronisation in low frequencies (red region below 10~Hz) accompanied by a desynchronisation in the beta band (blue region, 12-30~Hz), approximately 400~ms after arrow onset. These spectrotemporal patterns were consistently observed for saccades in all directions.
The spectrograms for the initial saccade (Figure~\ref{fig:spectrograms_main}(b)) showed activity resembling the P300 potentials observed in Figure~\ref{fig:spectrograms_main}(a), occurring prior to the saccade onset ($t=0$). This suggests that the initial saccade activity was confounded by overlapping P300 responses.
Moreover, Figure~\ref{fig:spectrograms_main}(c) did not reveal any distinct or consistent patterns for the back saccade.

For the Free-Viewing Task, Figures~\ref{fig:spectrograms_main}(d) and~\ref{fig:spectrograms_main}(e) show the spectrograms of IC7 for the initial and back saccades, respectively. Strong low-frequency synchronisation (below 15~Hz) was observed across most directions, indicated by red regions centred around the black vertical line ($t=0$) marking saccade onset. This pattern suggests the presence of consistent, time-locked spectral features associated with the initiation of free-viewing saccades.

An interesting observation was that certain saccade directions showed spectral features that were not present in other directions. For example, in the Visually-Guided Task, the `up' and `down' initial saccades (Figure~\ref{fig:spectrograms_main}(b)) showed beta-band (12-30~Hz) activity with a polarity reversal in spectral features around saccade onset: desynchronisation (blue) before the onset transitioned to synchronisation (red) afterwards. Similarly, in the Free-Viewing Task, the `up' initial saccade (Figure~\ref{fig:spectrograms_main}(d)) exhibited beta desynchronisation (blue regions) both before and after the saccade onset, which was not present in other directions. The presence of beta-band activity specifically for `up' saccades suggests the existence of direction-specific neural correlates that were not consistently observed across all directions.

\subsection{Saccade Classification}

As individual trials followed a bidirectional saccade paradigm, three categories of results are reported: (i)~decoding the `initial saccade', (ii)~decoding the `back saccade', and (iii)~decoding the `combined saccade' (initial and back saccades). The `initial' and `back' saccades were analysed separately, as the back saccade is a reflexive movement toward a predetermined target, whereas the initial saccade requires the user to select a direction, involving greater cognitive planning. This distinction was particularly important for the Visually-Guided Task, where initial saccades were preceded by a visual cue, potentially causing overlap between cue-related and saccade-related responses. In contrast, the back saccades were not confounded by such visual stimuli and therefore more directly reflected oculomotor planning.

\begin{figure}[!htbp]
    \centering
    \includegraphics[width=1\linewidth]{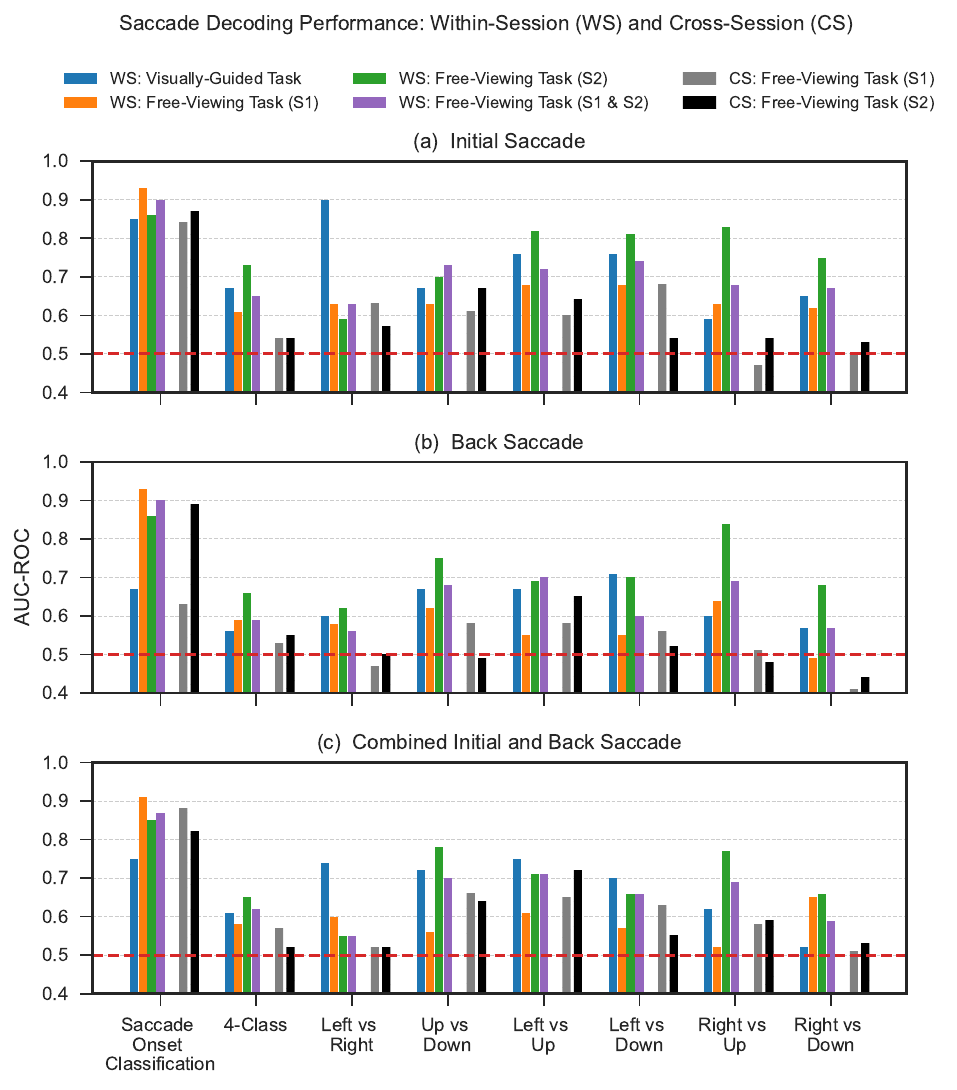}
    \caption[Saccade classification performance using sensor space data]{
    Saccade classification performance using sensor space data. Each panel presents AUC-ROC scores for different decoding tasks: (a)~initial saccades, (b)~back saccades, and (c)~combined initial and back saccades. The left-most group of bars in each subplot corresponds to saccade onset classification (fixation vs.\ saccade), while the remaining groups represent saccade direction classification results. Performance is reported for within-session~(WS) decoding on the Visually-Guided Task and Free-Viewing Task (Sessions~1 and~2), as well as for cross-session~(CS) decoding across sessions of the Free-Viewing Task. The red dashed line denotes chance-level performance (AUC~=~0.5).}
    \label{fig:saccade-classification-sensor-space}
\end{figure}

Figure~\ref{fig:saccade-classification-sensor-space} presents the results of decoding within a \mbox{[$-0.5$,~$0.5$]~s} interval around saccade onset using sensor space data.  Panels~(a), (b), and~(c) display decoding performance for initial saccades, back saccades, and combined saccades, respectively. Performance was measured using AUC-ROC scores, with the red horizontal dashed line indicating chance-level performance (AUC~=~0.5). The bar colours represented different task types: blue for within-session~(WS) decoding of the Visually-Guided Task, and orange, green, and purple for Session~1, Session~2, and pooled sessions of the Free-Viewing Task. Cross-session~(CS) decoding results are shown in grey and black: grey bars indicate training on Session~2 and testing on Session~1, while black bars indicate the reverse. Each subplot includes decoding results for both saccade onset classification (fixation vs.\ saccade) and saccade direction classification (4-class and pairwise binary), grouped separately and labelled along the x-axis.

The `Saccade Onset Classification' (left-most group of bar plots) emerged as a comparatively easier task, with AUC scores exceeding 0.80 in most cases across both within-session and cross-session decoding settings, and across all saccade types (initial, back, and combined). This strong performance was likely due to the direction-independent nature of saccadic ERP features. For the Visually-Guided Task, decoding initial saccades against fixation trials (blue bars in the left-most group in Figure~\ref{fig:saccade-classification-sensor-space}(a)) yielded the highest scores. However, performance declined with the inclusion of back saccades (blue bars in Figure~\ref{fig:saccade-classification-sensor-space}(b)-(c)). In contrast, within-session performance for the Free-Viewing Task (orange, green, and purple bars) remained consistently high, with AUC scores exceeding 0.85 across both sessions and all saccade types. Even under cross-session evaluation (grey and black bars), AUC scores generally remained above 0.80, except in the case of back saccades (Figure~\ref{fig:saccade-classification-sensor-space}(b)), where training on Session~2 and testing on Session~1 resulted in a reduced AUC of 0.63.

On the other hand, `saccade direction classification' was considerably more challenging, with decoding performance notably higher for initial saccades (Figure~\ref{fig:saccade-classification-sensor-space}(a)) than for back saccades (Figure~\ref{fig:saccade-classification-sensor-space}(b)). This was likely due to the planning involved in the preparation of the initial saccade, which also explains why previous oculomotor BCI studies focused exclusively on decoding this movement. Among the classification tasks, 4-class direction decoding proved the most difficult, while pairwise binary classification generally yielded better results. For within-session decoding, in most cases, Session~2 of the Free-Viewing Task (green bars) outperformed both Session~1 (orange) and the pooled data (purple), suggesting higher signal quality or engagement during that session. In contrast, cross-session direction classification was the most difficult setting, with AUC scores often falling below chance, particularly in the back saccade condition (Figure~\ref{fig:saccade-classification-sensor-space}(b)).

For reference, Figure~\ref{fig:saccade-classification-source-space} in \ref{Appendix-Decoding-ICs} shows the source space decoding results, which were consistent with the sensor space trends. Source space decoding generally achieved similar or higher AUC scores for both onset and direction classification. Notably, for cross-session saccade onset classification, AUC scores exceeded 0.85 for all conditions (initial, back and combined saccade), indicating strong generalisation.

\paragraph{Decoding using Pre-Saccade Data}

We also assessed the specific contribution of the pre-saccade interval to overall decoding performance. This distinction is important because decoding based on post-saccade data is more likely to reflect activity related to eye muscle movements, whereas oculomotor BCIs aim to decode the neural correlates of saccade planning. To investigate this, decoding was performed separately on pre-saccade, post-saccade, and combined pre+post-saccadic neural activity, with the results for decoding source space data presented in Table~\ref{tab:DecodingResultsPrevsPost}. Unlike previous analyses, a shorter epoch duration of 0.5~s was used instead of 1~s to prevent overlap between pre-saccade and post-saccade intervals of adjacent saccade events in the Free-Viewing Task. The results show that decoding performance using the pre-saccade interval is comparable to, though slightly lower than, that achieved with the post-saccade interval or the combined pre+post interval.

\input{Table1_Decoding_PreVsPost_Saccade}

%% file: Table1_Decoding_PreVsPost_Saccade.tex

\begin{table}[!htbp]
\renewcommand{\arraystretch}{1.1} 
\small 
\caption[Pre-saccade versus post-saccade saccade classification]{Comparison of AUC-ROC scores for decoding three different 500~ms intervals relative to initial saccade onset using sensor space data. 
The `pre+post', `pre', and `post' intervals correspond to the time windows \([-250, 250]\)~ms, \([-500, 0]\)~ms, and \([0, 500]\)~ms, respectively, all time-locked to saccade onset ($t=0$). 
`S1', `S2' and `S1+S2' refer to data from session~1, session~2, and the pooled data, respectively, of the Free-Viewing Task.
`Test: Free-Viewing S1' indicates training on Session~2 and testing on Session~1, while `Test: Free-Viewing S2' refers to training on Session~1 and testing on Session~2.}
\begin{tabularx}{\textwidth}{@{}p{3.7cm}XXXXXXXXX@{}} 

\toprule

\hline
\textbf{Task} & \multicolumn{3}{c}{\textbf{Saccade Onset}} & \multicolumn{3}{c}{\textbf{Left vs. Up}} & \multicolumn{3}{c}{\textbf{Left vs. Down}} \\
\hline
Saccade Interval & Pre+\newline Post & Pre & Post &  Pre+\newline Post & Pre & Post &  Pre+\newline Post & Pre & Post \\
\hline

\midrule
\multicolumn{9}{c}{\textit{Within-Session Testing}} \\ 

Visually-Guided      & 0.90 & 0.88 & 0.81 & 0.76 & 0.75 & 0.75 & 0.75 & 0.71 & 0.73 \\
Free-Viewing S1    & 0.89 & 0.86 & 0.91 & 0.68 & 0.67 & 0.69 & 0.69 & 0.70 & 0.67 \\
Free-Viewing S1   & 0.80 & 0.75 & 0.85 & 0.82 & 0.82 & 0.82 & 0.81 & 0.79 & 0.77 \\
Free-Viewing S1+S2 & 0.87 & 0.82 & 0.88 & 0.72 & 0.72 & 0.73 & 0.74 & 0.75 & 0.75 \\
\\
\multicolumn{9}{c}{\textit{Cross-Session Testing}} \\ 
Test: Free-Viewing S1  & 0.86 & 0.81 & 0.85 & 0.61 & 0.60 & 0.62 & 0.67 & 0.68 & 0.70 \\
Test: Free-Viewing S2 & 0.80 & 0.78 & 0.83 & 0.70 & 0.67 & 0.57 & 0.64 & 0.61 & 0.67 \\

\hline
\bottomrule 
\end{tabularx}

\label{tab:DecodingResultsPrevsPost}
\end{table}

%% file: 4_Discussion.tex
\section{Discussion}

This study demonstrates that neural correlates of saccades are present in brain signals recorded by the Stentrode and that saccade trials can be reliably distinguished from fixation trials. In the time domain, ERP analysis revealed saccadic potentials that emerge before the saccade onset and peak after saccade execution. These potentials are also visible at the individual trial level, particularly for free-viewing saccades. In the frequency domain, saccade trials are more distinguishable from fixation trials in low frequency bands (below 30~Hz), highlighting the spectral features relevant for decoding saccadic activity. These findings confirm that saccadic eye movements evoke distinct neural responses, which can be utilised for BCI applications.

\subsection{Comparison of Saccadic Potentials between Visually-Guided and Free-Viewing Saccade Tasks}
\label{ssec:wait-times}
\begin{figure}[!tb]
    \centering
    \includegraphics[width=0.6\linewidth]{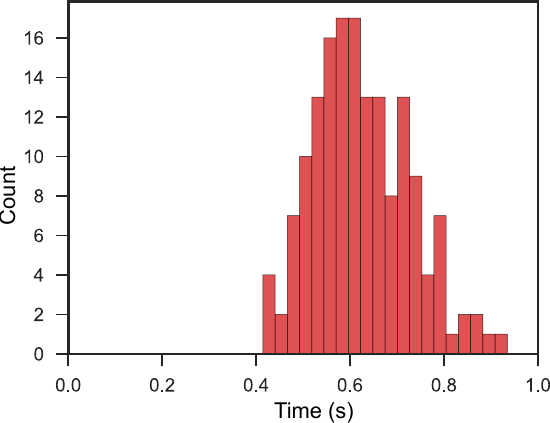}
    \caption{Distribution of wait times between the appearance of the visual cue and the actual saccade onset in the Visually-Guided Saccade Task. On average, the participant initiated a saccade 626~ms after the arrow appeared, with a minimum latency of 414~ms.}
    \label{fig:wait_times_vgt}
\end{figure}

While the evoked responses of free-viewing saccades time-locked to saccade onset showed identical saccade-related potentials for each saccade direction (left, right, up, and down), they had a different time course than the visually-guided saccades, which was particularly true for the initial saccade in the Visually-Guided Task.
This can be explained by the presence of P300 potentials in the Visually-Guided Task due to the appearance of visual cues (arrows onsets) that were absent in the Free-Viewing Task. These evoked responses overlap with individual trials differently; therefore, ERPs appeared differently in different directions. The diverse impact of P300 potentials on individual trials can be explained by investigating the distribution of wait times between the onset of the cue (arrow) and the actual onset of the saccade. As shown in Figure~\ref{fig:wait_times_vgt}, for valid visually-guided trials ($N~=~160$), the user moved their eyes on average 626~ms after arrow appearance (min~=~414~ms, max~=~935~ms, median~=~612~ms, std~=~103~ms). This explains why average ERPs appeared differently in different directions due to the impact of P300 potentials. It appeared that saccade-related ERPs follow cue-related ERPs, since P300 occurs approximately 300~ms after visual stimulus onset and the distribution of wait times suggests that their effect should be reduced around the onset of the saccades. However, it is difficult to separate saccadic potentials from P300 for the Visually-Guided Task as both features overlap in the time domain at low frequencies.

\subsection{Potential for an Endovascular P300 BCI}
As discussed in Section~\ref{sssec:p300}, the presence of visual cues preceding visually-guided saccades evoked a cue-related P300 potential that peaked approximately 300-400~ms after the appearance of the arrow. Although noisy, similar evoked responses were observed in different saccade directions, even at the level of individual trials. This P300 signal has been widely used in BCI applications, particularly in P300 spellers~\cite{philip2020visual}, where the user is asked to focus on a screen where various targets are flashed in a sequence following an oddball paradigm. The target stimulus to which the user is attending elicits a stronger P300 response, allowing the system to identify the intended selection. Previous research has shown that such P300 potentials can also be elicited using auditory or tactile stimuli~\cite{pan2022advances},and Guger et al.~\cite{guger2017complete} demonstrated that two out of three completely locked-in patients in their study were able to communicate using a vibrotactile P300 BCI system. However, a key limitation of P300-based BCIs is that they rely on exogenous stimulation, requiring users to attend to repetitive stimuli presented at fixed intervals, which can lead to fatigue with prolonged use~\cite{peters2020ssvep}. In contrast, endogenous BCI paradigms, such as motor imagery or oculomotor control, allow users to initiate control voluntarily, without dependence on external cues, making them more suitable for long-term use.


\subsection{Neural Correlates or EOG Artefacts?}
An important question is whether the proposed oculomotor BCI system decodes neural activity related to the planning of saccades, or whether it merely detects EOG artefacts that are commonly observed in scalp EEG recordings~\cite{ranjan2021ocular}. 
We present several reasons to support the interpretation that the saccade-related potentials observed in our study are neural in origin and cannot be attributed to EOG artefacts.

First, visual inspection of both the original sensor space data and the source space data (after applying ICA) did not reveal periodic waveforms typically associated with EOG artefacts in EEG.
Second, the Stentrode is placed in the superior sagittal sinus, which is anatomically distant from the corneo-retinal dipole. This location is more analogous to subdural ECoG arrays, which are generally unaffected by EOG artefacts.
Third, the saccade-related potentials shown in Figure~\ref{fig:ERP_Main}(e--f) include a prominent pre-saccade component that precedes physical eye movement. Since EOG artefacts arise from actual eye motion, they cannot account for this pre-movement activity.
Fourth, EOG artefacts in EEG typically show polarity reversal across contralateral directions and differ substantially between horizontal and vertical saccades~\cite{heo2017novel, plochl2012combining, romero2009ocular, jia2019measurement}. In contrast, the saccade-related potentials recorded in our study appear consistent across all directions, suggesting a common underlying neural process rather than direction-dependent EOG artefacts.
Finally, the strongest evidence comes from our decoding analysis. As shown in Table~\ref{tab:DecodingResultsPrevsPost}, the classification performance using only pre-saccade data is comparable to performance using both pre-saccade and post-saccade windows. Since EOG artefacts cannot precede eye movement onset, this indicates that the classifier relies on pre-movement neural activity rather than ocular artefacts.

\subsection{Comparison with other Oculomotor BCI Studies}

\subsubsection{Pre-Saccade Potentials:}
According to Jagla et al.~\cite{jagla2007saccadic}, pre-saccadic eye movement-related potentials can be categorised into three components: (i)~Pre-Motion Negativity, which begins 1-3~s before saccade onset and may include readiness potentials, (ii)~Pre-Motion Positivity, which starts 100-150~ms before saccade onset and is associated with motor program formulation, and (iii)~the Spike Potential, which emerges within 100~ms before saccade execution and peaks just before or after saccade onset. 
The exact amplitude, duration, and topography of these pre-saccade potentials vary between studies~\cite{thickbroom1985cerebral, klostermann1994presaccadic} due to differences in experimental protocols.

In this study, we prioritised the real-time nature of practical BCI tasks, which resulted in inter-saccade intervals too short to observe pre-saccade negativity or readiness potentials. The saccadic potentials of free-viewing saccades (see Figure~\ref{fig:ERP_Main}(e-f)), revealed a signal that emerged approximately 50~ms before saccade onset, peaked around 50~ms after saccade execution, and lasted approximately from \([-50, 200]\)~ms relative to saccade onset. This timing aligns with the general description of saccade potentials, which typically begin within 100~ms before saccade onset and peak around the onset.  The exact timing of pre-saccadic Spike Potentials (SPs) varies across studies. For example, Boylan et al.~\cite{boylan1988presaccadic} reported that, for horizontal saccades, SPs begin 12-4.4~ms before saccade onset, with peaks occurring between saccade onset and 7.8~ms post-saccade. Similarly, Doig et al.~\cite{doig1990presaccadic} found that, for vertical saccades, SPs begin 17.3-6.8~ms before saccade onset, peaking between 3.4~ms pre-saccade and 5.2~ms post-saccade. Furthermore, Funase et al. demonstrated the existence of pre-saccade potentials across various saccade types, including visually-guided and auditory-guided saccades~\cite{funase2002analysis, funase2007single}, memory-guided saccades~\cite{ohno2007analysis}, and free-viewing saccades~\cite{funase2011research}, with peak pre-saccade potentials occurring within 15~ms before saccade onset. Thus, the timing of saccade-related potentials observed for free-viewing saccades in our study is consistent with previous research.

\subsubsection{Saccade Classification:}
Although the results of the saccade onset classification reported in this study are high, the saccade direction classification does not outperform the results of the invasive oculomotor studies in nonhuman primates (discussed in the Introduction (Section~\ref{Introduction})). The ECoG-based oculomotor BCI study~\cite{lee2017decoding} reported mean accuracies of 80\% for two directions and 43\% for four directions. For comparison, we evaluated the accuracy of our best-performing pipelines and achieved the following mean accuracies: saccade onset classification, 82\% (chance 50\%); four-class classification, 44\% (chance 28\%); left vs.\ up saccade, 67\% (chance 52\%); and left vs.\ down saccade, 76\% (chance 55\%). These values represent mean within-session decoding scores across three sessions (one session of Visually-Guided and two sessions of the Free-Viewing Task). The reported chance accuracies reflect the slight imbalances that existed in the number of trials for each direction, caused by the self-paced nature of the experiment and the removal of bad trials. Although the four-class decoding results are comparable, the ECoG study still outperformed our binary saccade direction classification results. However, considering that 80\% is a minimum device accuracy benchmark for BCI usability~\cite{brannigan2024brain}, our system performed well for saccade onset classification. Furthermore, Ohmae et al.~\cite{ohmae2015decoding} reported similar findings, demonstrating that distinguishing saccades from fixations is a relatively easier task compared to classifying saccade directions. Their study showed that the best performance was achieved approximately 200~ms after saccade onset. This observation aligns with the saccadic potentials observed in IC7 for the Free-Viewing Task, which persist for about 200~ms after saccade onset, as illustrated in Figure~\ref{fig:ERP_Main}(e-f).

The primary reason for the performance difference between our study and other invasive oculomotor studies in macaques lies in the type of neural implant and the resolution of the recorded neural signals. Invasive studies used microelectrode arrays capable of recording single-cell activity, whereas the Stentrode can capture only broadband signals. Furthermore, microelectrodes were implanted in closer proximity to key oculomotor regions, such as the Frontal Eye Field~(FEF), Supplementary Eye Field~(SEF), dorsolateral prefrontal cortex~(DLPFC), and Lateral Intraparietal area~(LIP), with some studies utilising multiple implants to record from multiple sites simultaneously. In contrast, in our study, the recordings were made from a single Stentrode device implanted near the Supplementary Motor Area~(SMA) to capture motor cortex activity. Although the SMA is near the SEF, the recorded signals may include substantial neural activity not related to saccade planning.

Furthermore, cross-session decoding performance was lower compared to within-session classification. Decoding across sessions, particularly when data are collected on different days, is a more challenging BCI task due to neural signal non-stationarity, environmental variability, and changes in the user's cognitive and physiological state. In our study, data collection was limited to two sessions for the Free-Viewing Task and one session for the Visually-Guided Task, due to the participant's declining health. In contrast, developing a generalisable and robust cross-session classifier typically requires data from multiple sessions across different days. To mitigate this limitation, data augmentation techniques can be employed to artificially increase the effective size of the training set~\cite{rommel2022data}. In addition to collecting more data in future studies, methods such as domain adaptation~\cite{wu2016online} and transfer learning~\cite{wu2022transfer} could also be used to improve generalisation across sessions.

\subsubsection{Experiment Design Considerations:}
In terms of experimental design, a key difference in our approach was the use of self-paced saccades, where the user was instructed to wait briefly before executing a saccade, in order to extend the saccade planning interval. This differs from classical delayed visually-guided or memory-guided saccade tasks, in which the wait time before saccade execution is controlled by the experimental paradigm using external cues. Moreover, both previous oculomotor studies that successfully decoded saccades using only pre-saccade data~\cite{jia2017decoding, graf2014brain} employed memory-guided saccades. However, we chose not to adopt this approach, as saccade planning features could become confounded by memory and attention processes. Additionally, after identifying the presence of P300 potentials elicited by the target cue in the Visually-Guided Task, we developed a fully self-paced Free-Viewing Task, which has not been explored in previous invasive oculomotor BCI studies. This task allowed the user to make saccades at their own pace without external cues. To naturally introduce a short planning interval, the user was instructed to wait briefly before performing each saccade.

The Visually-Guided Task and Free-Viewing Task differ in the neural features associated with saccade planning. Bruce et al.~\cite{bruce1985primate} demonstrated that a Visually-Guided Task activates three types of neurons: visual, visuomovement, and movement neurons. In contrast, free-viewing saccades occur without an external cue, engaging only movement neurons. This difference suggests that the Visually-Guided Task should elicit stronger neural signals, which may explain why previous invasive oculomotor studies predominantly used the Visually-Guided Task. For instance, Ohmae et al.~\cite{ohmae2015decoding}~found that instead of relying solely on motor neurons, the performance of oculomotor BCI increased by incorporating all types of neurons that activate between stimulus presentation and saccade execution. Similarly,  Tauscher et al.~\cite{tauscher2018analysis} compared the Visually-Guided and Free-Viewing Tasks in human participants, reporting more pronounced ERP responses time-locked to saccade onset in the Free-Viewing Task. However, confirming this distinction in our study is challenging, as the low-frequency ERP features of saccades overlap with the P300 response of visual cue in the Visually-Guided Task.

\subsubsection{Data Acquisition Setup:}
The choice of data collection setup may have influenced task performance. We prioritised real-world applicability by incorporating hardware already used by the participant for everyday digital access.

One such consideration was the size of the monitor. Although we used a 15-inch laptop, previous oculomotor BCI studies have generally used larger monitors, such as 19''~\cite{Reza2014EEGTRACKING}, 24''~\cite{drouin2016using}, and 27''~\cite{dietrich2017towards}. Larger monitors allow for an increased spacing between targets on the screen, thereby generating larger visual angles. Belkacem et al.~\cite{belkacem2014classification} demonstrated that increasing the visual angle from 5$^\circ$ to 10$^\circ$, 20$^\circ$, and 30$^\circ$ correspondingly increased the classification accuracy from 66.5\% to 73.2\%, 81.2\%, and 86.8\% respectively. These results suggest that larger visual angles produce stronger saccade-evoked potentials, thereby improving oculomotor BCI classification performance. With a spacing of approximately 8.5~cm between the fixation centre and the targets, and a viewing distance of 50~cm, we obtained a visual angle of approximately 5$^\circ$. Although this choice highlights the feasibility of deploying oculomotor BCIs in realistic home environments, it may have limited the amplitude of the saccades and consequently reduced the strength of the associated evoked responses.

Similar practical considerations influenced the selection of eye-tracking hardware. The participant was already using the Tobii PCEye eye tracker as part of his existing assistive setup. While using the same device facilitated seamless integration into the participant's daily routine, it also imposed certain limitations, notably a lower sampling rate and restricted access to raw gaze data. Specifically, the Tobii PCEye operates at a sampling rate of 33~Hz and does not offer access to raw gaze data via an open-access API. Consequently, gaze positions were indirectly inferred by using cursor positions on the screen. In the Visually-Guided Task, cursor positions were constrained by the 30~Hz frame rate of the front-facing camera, whereas in the Free-Viewing Task, they were limited by the 60~Hz refresh rate of the screen, as camera data were not recorded in the latter. These relatively low sampling rates correspond to temporal resolutions of approximately 33.3~ms and 16.7~ms, respectively, potentially introducing small temporal misalignment (jitter) between neural signals and eye-movement events. This temporal jitter may have reduced the precision of time-locked ERP features, highlighting a broader issue in multimodal BCI research where high-frequency neural recordings (e.g., 2000~Hz) must be temporally aligned with lower-frequency signals from secondary devices like eye trackers, video cameras, or electromyography systems. Although research-grade eye trackers, such as SR Research’s EyeLink 1000 with a 2000~Hz sampling rate, offer superior temporal precision and automated differentiation between saccades, fixations, blinks and microsaccades, our main objective was to demonstrate the feasibility of practical BCI decoding. Despite its limitations, the Tobii PCEye provided sufficient temporal resolution to accurately label saccade onset times in the current study.

\subsection{User-Centred and Translational Considerations for Home-Based BCI}
The performance of BCI systems can be influenced by various factors beyond purely technical considerations, including user engagement, cognitive fatigue, and the practical realities of the deployment environment. In the late stages of ALS, patients often experience diminished capacity for sustained attention, which affects their ability to participate in BCI tasks. During data collection, it was observed that the participant reported feeling drowsy during the Visually-Guided Task and suggested improvements to increase engagement. In response, additional rest breaks were incorporated within each data collection run, and the Free-Viewing Task interface was refined to facilitate a better focus on individual targets. These adjustments appeared to improve participant relaxation and sustained attention.

Furthermore, this study was conducted at the participant's home instead of a controlled laboratory setting, introducing practical and environmental challenges. During data collection, several uncontrolled factors were observed, such as noise from nearby construction and distractions from household pets. Although these factors increased variability and noise in the data, they also highlight the translational relevance of the findings.


\subsection{Potential for Completely Locked-in Syndrome Patients?}
Our study involved overt eye movements, but previous research suggests that these may not be essential for effective BCI operation. This implies that such systems could help patients with completely locked-in syndrome who have lost control over voluntary eye movements. For instance, Jia et al.~\cite{jia2017decoding}~trained monkeys in a memory-guided saccade task and successfully decoded target selection from the pre-saccade memory interval. After some training, one of the monkeys was able to control the system without overt eye movements, demonstrating the feasibility of a `covert oculomotor BCI'. Similarly, Graf et al.~\cite{graf2014brain} predicted saccade plans in monkeys based on brain signals that correspond to covert eye movements. These findings suggest that oculomotor BCIs could potentially benefit patients with complete paralysis. This concept parallels motor imagery BCIs, where imagined limb movements generate neural signals similar to actual movements~\cite{mustile2024neural}, allowing BCIs for individuals with paralysis. However, further testing with a patient in a completely locked-in state is required to assess the full potential of a covert or imagined oculomotor BCI.

\subsection{Impact and Utility of Oculomotor BCIs}
An oculomotor BCI could enable 2D cursor control similar to an eye tracker, as the eye gaze directly maps to the corresponding 2D cursor positions. Although typical eye-tracking software requires users to maintain fixation for a period to select a character, typing speed can be increased by using swipe typing~\cite{mifsud2023hmm}, which can be combined with language correction models to reduce errors. These strategies suggest the feasibility of oculomotor BCIs for continuous cursor control. However, current studies have mainly demonstrated discrete control, making them more suitable for `end-point decoding'.

The goal-oriented nature of saccades makes them particularly well-suited for tasks involving rapid serial selections from a set of spatially distributed targets. A potential application is augmentative and alternative communication (AAC) software, which typically involves selecting from a set of possible words or options displayed on the screen. For AAC applications, an oculomotor BCI offers certain advantages over conventional motor BCIs. For example, in a four-class selection task, a classical motor BCI would map movements of the right hand, left hand, both feet and tongue to the left, right, up and down selections, which is an indirect mapping. In contrast, a four-class oculomotor control system would use left, right, up, and down saccades, which are more intuitive and much faster. However, we acknowledge that the current results of the oculomotor BCI decoding presented here are less accurate compared to the well-established motor decoding BCIs~\cite{patrick2025state}. Further research with human subjects may help achieve performance comparable to that of invasive oculomotor studies in monkeys.

In terms of experiment design, the Visually-Guided Saccade Paradigm can be used to implement a synchronous exogenous BCI system, where a visual cue prompts the user before performing eye movements, allowing them to make selections through those movements. Exogenous cue-based P300~\cite{philip2020visual} or SSVEP~\cite{besharat2024comparative} BCIs already serve this purpose, and, as observed, strong P300 signals can be obtained through Stentrode recordings with a simple visual cue (without flashes). However, for long-term practical use, endogenous BCI control is necessary, which does not require external cues. Free-viewing saccades offer a more natural, asynchronous endogenous BCI control, where the decoder continuously looks for potential eye movement patterns, registering the onset of a saccade and using its direction to make a selection. However, this realistic approach comes at a cost: the proposed self-paced free-viewing saccade is more challenging than the previous scenarios, as it (i)~does not activate visual or visuomovement neurons like visually-guided sacccades, and (ii)~lacks a dedicated pre-saccade `memory' phase, making the planning interval smaller and increasing the task complexity.

An oculomotor BCI signal can be used to control external devices such as robotic arms or wheelchairs. For example, Aziz et al.~\cite{aziz2014hmm} used saccade-related features from EEG data for wheelchair control. Additionally, an oculomotor BCI can serve as a complementary control channel alongside conventional BCI paradigms, allowing simultaneous signal acquisition without requiring additional obtrusive hardware (e.g., EOG electrodes) or expensive equipment (e.g., eye trackers). For example, Samadi et al.~\cite{samadi2014eeg} integrated EOG features from EEG with an SSVEP-based EEG paradigm for eye tracking, achieving 89\% accuracy in predicting eight-direction saccades. This highlights the potential of oculomotor BCIs both as standalone systems and as components of multimodal BCI frameworks.


%% file: 5_Conclusion.tex


\section{Conclusion}

This proof-of-concept study demonstrates the potential of an oculomotor BCI using a minimally invasive endovascular Stentrode device in a human user. Our findings confirm the presence of neural correlates associated with saccadic eye movements and highlight the feasibility of decoding saccade-related brain activity for assistive BCI applications.

Although this study marks an important step, more research is needed to validate these findings over large number of sessions and in multiple participants. Furthermore, while the location of the Stentrode implant in the current patient was optimised for motor decoding, stronger oculomotor control signals may require placement closer to oculomotor brain regions and may also involve recording from multiple brain regions.
Implementing an asynchronous online saccade onset detection system and improving the performance of saccade direction classification will enable the development of a two-stage real-time oculomotor BCI, where the first stage detects saccade onset, followed by classification of the intended direction. Advancing these aspects will help establish oculomotor BCIs as a practical communication tool for individuals with severe motor impairments.

%% file: Acknowledgment.tex
\section*{Acknowledgment}
This research was conducted in the Australian Research Council Training Centre in
Cognitive Computing for Medical Technologies (Project Number~ICI70200030), funded by the Australian Government. We thank the participant for providing consent to take part in this study and Z.~Faraz for his assistance in collecting patient data. We thank all members of the ARC Training Centre and Synchron Inc.\ for their valuable feedback and discussions. We acknowledge the use of a large language model (ChatGPT, OpenAI) to assist in improving grammar and clarity, and the authors take full responsibility for the content of this manuscript.

%% file: 6_Appendices.tex
\newpage
\appendix


\section{Comparison of Classification Algorithms}
\label{Appendix-Comparison-Classifiers}

\begin{figure}[!htbp]
    \centering
    \includegraphics[width=1\linewidth]{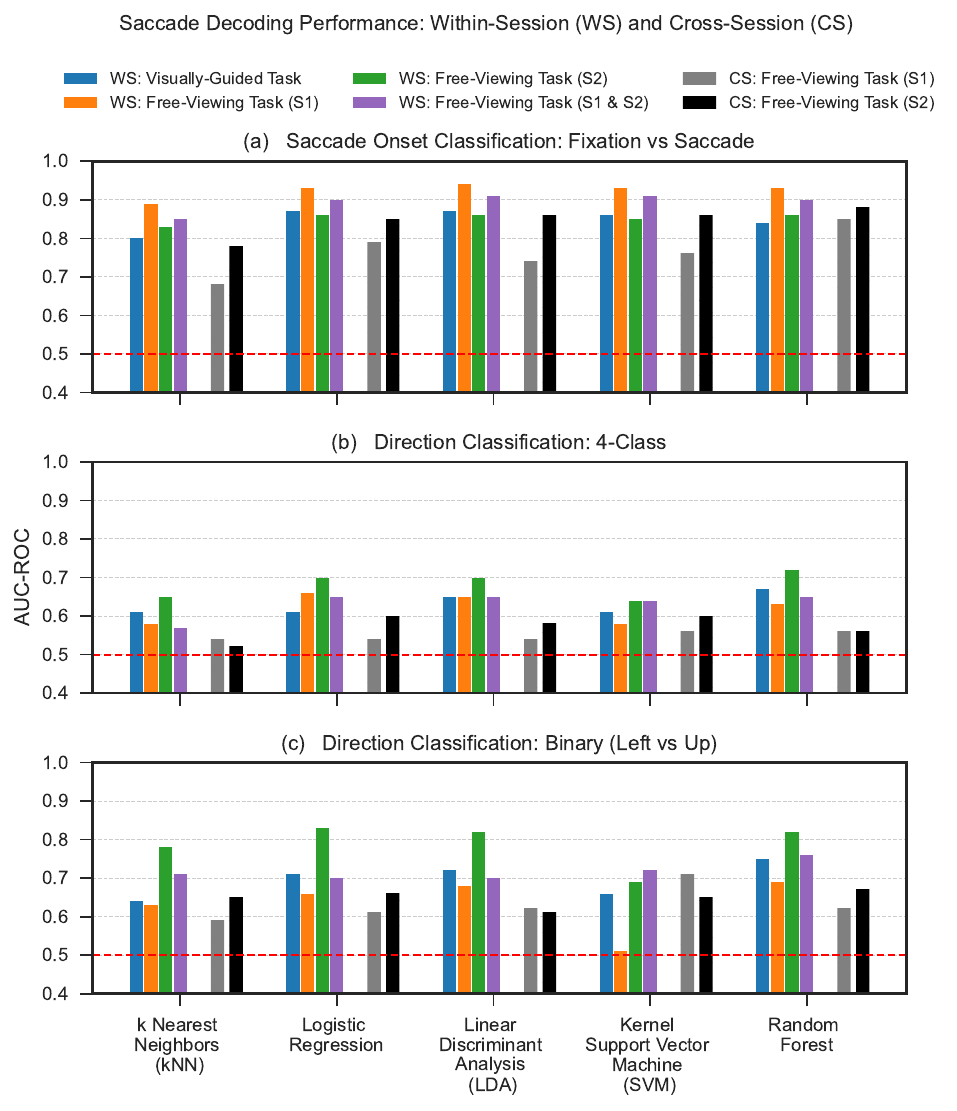}
    \caption{
    Comparison of classification performance across different machine learning algorithms (x-axis) using sensor space data. Each panel shows AUC-ROC scores for a specific decoding task: 
    (a)~Saccade onset classification (fixation vs.\ saccade), 
    (b)~Four-class saccade direction classification, and 
    (c)~Binary direction classification (left vs.\ up saccade). 
    Performance is reported for within-session (WS) decoding on both the Visually-Guided and Free-Viewing Tasks (Sessions 1 and 2), while cross-session (CS) decoding is reported for the Free-Viewing Task. The red dashed line indicates chance-level performance (AUC = 0.5).}
    \label{fig:saccade-classification-comparison-classifiers}
\end{figure}

To select the most suitable classification algorithm, we compared the performance of five widely used methods: $k$-Nearest Neighbors (kNN), Logistic Regression, Linear Discriminant Analysis (LDA), Support Vector Machines (SVM) with an RBF kernel, and Random Forest. Figure~\ref{fig:saccade-classification-comparison-classifiers} illustrates the classification performance of these algorithms using mean AUC-ROC scores. Each set of bar plots represents within-session (WS) and cross-session (CS) decoding performance for each classifier (x-axis). Panel (a) displays the results for saccade onset classification (fixation vs. saccade), while panels (b) and (c) show the performance for saccade direction classification tasks in four-class and binary (left vs. down) scenarios, respectively. The binary left vs. down comparison was specifically selected because it had  demonstrated superior performance compared to other binary classification pairings in the main results.

Overall, performance across algorithms was comparable, with the Random Forest classifier performing better in certain cases, such as cross-session decoding for saccade onset classification (panel (a)). Since no single algorithm consistently outperformed others across all tasks, we calculated the average classification performance (mean $\pm$ standard deviation) for each algorithm. As shown in Table~\ref{tab:Decoding_Comparison_Classifiers}, Random Forest slightly outperformed other algorithms in both the saccade onset and binary direction (left vs. down) classification tasks.


\input{TableA1_Decoding_Comparison_Classifiers}

\newpage
\section{Saccadic Potentials}
\label{Appendix-Saccadic-Potentials}

\begin{figure}[!htbp]
    \centering
    \includegraphics[width=1\linewidth]{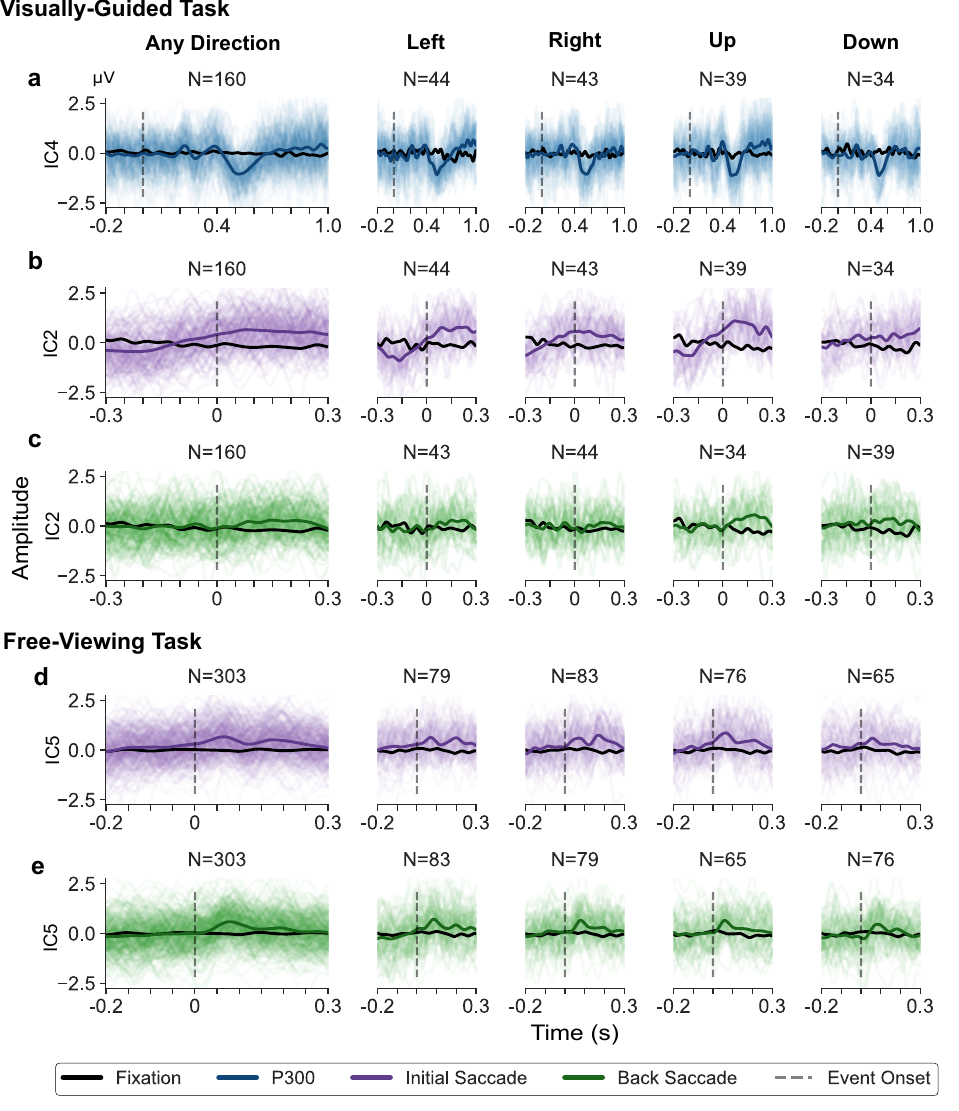}
    \caption{Evoked responses for the Visually-Guided (a-c) and Free-Viewing Tasks (d-e), shown across directions (columns) and different events (rows). 
    (a) P300 potentials time-locked to cue onset ($t=0$) using IC4. 
    (b–f) Saccade-related ERPs time-locked to initial (b, d) and back (c, e) saccades using IC2 (Visually-Guided) and IC5 (Free-Viewing). 
    For details on labelling conventions, refer to the caption of figure~\ref{fig:ERP_Main}.}
    \label{fig:ERP_Appendix}
\end{figure}

\newpage
\section{Decoding using Source Space Data}
\label{Appendix-Decoding-ICs}

\begin{figure}[!htbp]
    \centering
    \includegraphics[width=1\linewidth]{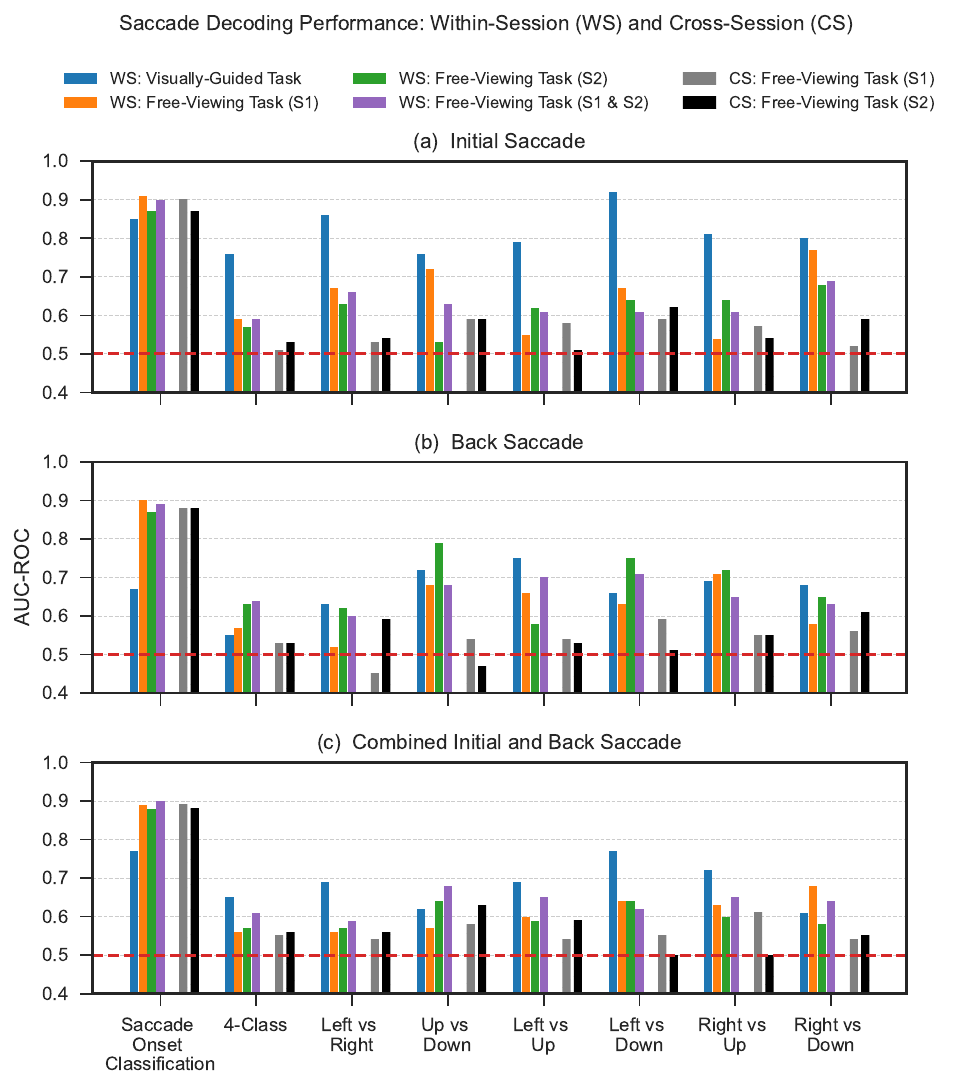}
    \caption[Saccade classification performance using source space data]{
    Saccade classification performance using source space data. Each panel presents AUC-ROC scores for different decoding tasks: (a)~initial saccades, (b)~back saccades, and (c)~combined initial and back saccades. 
    Performance is reported for within-session (WS) decoding on both the Visually-Guided and Free-Viewing Tasks (Sessions 1 and 2), while cross-session (CS) decoding is reported for the Free-Viewing Task. The red dashed line indicates chance-level performance (AUC = 0.5).}
    \label{fig:saccade-classification-source-space}
\end{figure}

%% file: TableA1_Decoding_Comparison_Classifiers.tex

\begin{table}[!htbp]
\renewcommand{\arraystretch}{1.2}
\small
\caption{
Classification performance (mean~$\pm$~standard deviation of AUC-ROC scores) for five classifiers across three tasks. Scores are averaged across within-session and cross-session evaluations, and across both Visually-Guided and Free-Viewing paradigms. Bolded results indicate the best-performing classifier for each task.}
\resizebox{\textwidth}{!}{%
\begin{tabular}{|c|ccc|}
\hline
\multirow{3}{*}{\textbf{\begin{tabular}[c]{@{}c@{}}Classification \\ Algorithm\end{tabular}}} &
  \multicolumn{3}{c|}{\textbf{Task}} \\ \cline{2-4} 
 &
  \multicolumn{1}{c|}{\multirow{2}{*}{\begin{tabular}[c]{@{}c@{}}Saccade \\ Onset\\ Classification\end{tabular}}} &
  \multicolumn{2}{c|}{\begin{tabular}[c]{@{}c@{}}Saccade Direction \\ Classification\end{tabular}} \\ \cline{3-4} 
 &
  \multicolumn{1}{c|}{} &
  \multicolumn{1}{c|}{4-Class} &
  \begin{tabular}[c]{@{}c@{}}Binary\\ (Left vs. Down)\end{tabular} \\ \hline
K Nearest Neighbours &
  \multicolumn{1}{c|}{0.81 ± 0.07} &
  \multicolumn{1}{c|}{0.58 ± 0.05} &
  0.67 ± 0.07 \\ \hline
Logistic Regression &
  \multicolumn{1}{c|}{0.87 ± 0.05} &
  \multicolumn{1}{c|}{\textbf{0.63 ± 0.06}} &
  0.70 ± 0.08 \\ \hline
Linear Discriminant Analysis &
  \multicolumn{1}{c|}{0.86 ± 0.07} &
  \multicolumn{1}{c|}{\textbf{0.63 ± 0.06}} &
  0.69 ± 0.08 \\ \hline
Kernel SVM &
  \multicolumn{1}{c|}{0.86 ± 0.06} &
  \multicolumn{1}{c|}{0.61 ± 0.03} &
  0.66 ± 0.08 \\ \hline
Random Forest &
  \multicolumn{1}{c|}{\textbf{0.88 ± 0.03}} &
  \multicolumn{1}{c|}{\textbf{0.63 ± 0.06}} &
  \textbf{0.72 ± 0.07} \\ \hline
\end{tabular}%
}
\label{tab:Decoding_Comparison_Classifiers}
\end{table}